\documentclass[superscriptaddress,aps,onecolumn,12pt,floatfix,altaffilletter,showpacs,showkeys,notitlepage,nofootinbib,preprintnumbers]{revtex4-2}

\usepackage[dvipdf,dvips,dvipdfmx]{graphicx}
\usepackage[colorlinks=true,citecolor=blue,linkcolor=blue]{hyperref}
\usepackage{amsmath}
\usepackage{bm}
\usepackage{amssymb}
\usepackage{epsfig}
\usepackage{epstopdf}
\usepackage{url}
\usepackage{color}
\usepackage[utf8]{inputenc} 
\usepackage{amsmath}
\usepackage{amssymb}
\usepackage{slashed}
\usepackage{cancel}
\usepackage{textcomp}
\usepackage{calc}
\usepackage{color}
\usepackage{soul,xcolor}
\allowdisplaybreaks

\catcode`\@=11
\def\lsim{\mathrel{\mathpalette\@versim<}}
\def\gsim{\mathrel{\mathpalette\@versim>}}
\def\@versim#1#2{\vcenter{\offinterlineskip
\ialign{$\m@th#1\hfil##\hfil$\crcr#2\crcr\sim\crcr } }}
\catcode`\@=12
\makeatletter
\def\lsim{\mathrel{\mathpalette\@versim<}}
\def\gsim{\mathrel{\mathpalette\@versim>}}
\def\@versim#1#2{\vcenter{\offinterlineskip
        \ialign{$\m@th#1\hfil##\hfil$\crcr#2\crcr\sim\crcr } }}
\makeatother
\newcommand{\vs}[1]{\vspace{#1 mm}}
\def\disp{\displaystyle}
\def\rmP{{\rm P}}
\def\rmT{{\rm T}}
\def\rmc{{\rm c}}

\def\bfx{{\boldsymbol x}}
\def\bfy{{\boldsymbol y}}
\def\bfp{{\boldsymbol p}}
\def\bfP{{\boldsymbol P}}
\def\bfq{{\boldsymbol q}}
\def\bfk{{\boldsymbol k}}

\def\Re{{\rm Re}\,}
\def\Im{{\rm Im}\,}

\def\calE{{\omega}}

\def\nn{\nonumber\\}

\def\what{\widehat}
\def\half{\frac12}
\def\sqr#1#2{{\vcenter{\hrule height.#2pt
      \hbox{\vrule width.#2pt height#1pt \kern#1pt
          \vrule width.#2pt}
      \hrule height.#2pt}}}
\def\bra#1{\left\langle{#1}\right|}
\def\ket#1{\left|{#1}\right\rangle}
\def\VEV#1{\left\langle{#1}\right\rangle}

\makeatletter

\begin{document}

\preprint
{}YITP-23-104


\medskip
\renewcommand{\thefootnote}{\fnsymbol{footnote}}

\begin{center}
{\Large\bf  
Unitarity Violation in Field Theories of \\
\vs{5} Lee-Wick's Complex Ghost}
\vs{20}

{\large
Jisuke Kubo,$^{1,2,}$\footnote{e-mail address: kubo@mpi-hd.mpg.de}
and Taichiro Kugo,$^{3,}$\footnote{e-mail address: kugo@yukawa.kyoto-u.ac.jp}
} \\
\vs{5}

$^1$
{\it 
Max-Planck-Institut f\"ur Kernphysik(MPIK), Saupfercheckweg 1, 69117 Heidelberg, Germany}

$^2$
{\it 
Department of Physics, University of Toyama, 3190 Gofuku, Toyama 930-8555, Japan
}

\vs{3}
$^3$
{\it Center for Gravitational Physics
and Quantum Information,
\\ Yukawa Institute for Theoretical Physics,
Kyoto University, Kyoto 606-8502, Japan}

\vs{15}
{\bf Abstract}
\end{center}

Theories with fourth-order derivatives, including the Lee-Wick
finite QED model and Quadratic Gravity, have 
a better UV behaviour, but the presence of 
negative metric ghost modes endanger  unitarity.
Noticing that the ghost acquires
a complex mass by radiative corrections, Lee and Wick, in particular,
claimed that such complex ghosts would never be created by collisions of
physical particles because of energy conservation, so that the
physical S-matrix unitarity must hold. 

We investigate the unitarity problem faithfully working in the
operator formalism of quantum field theory. 
When complex ghosts participate, a complex delta function
(generalization of Dirac delta function)
appears at each interaction vertex, which enforces a specific 
conservation law of complex energy. 
Its particular property
implies that the  naive Feynman rule is wrong if the four-momenta
are assigned to the internal lines after taking account of the conservation 
law in advance.
We show that the complex ghosts are actually created and unitarity is 
violated in such fourth-order derivative theories.
We also find a definite energy threshold below which the ghosts cannot be
created: The theories are unitary  and renormalizable below the threshold.

\renewcommand{\thefootnote}{\arabic{footnote}}
\setcounter{footnote}{0}

\section{Introduction}

It has long been a challenge  to construct a consistent quantum filed theory
(QFT) of gravity, though
a well-known candidate is Quadratic Gravity (QG) whose renormalizability
was proven by Stelle \cite{Stelle:1976gc}  long  ago.
The action of QG contains in addition to the Einstein-Hilbert term
$R$, the Weyl tensor squared $C_{\mu\nu\rho\sigma}^2$
as well as $R^2$ (see e.g. \cite{Mannheim:2011ds,Salvio:2018crh} for a review). The canonical dimension of these two terms is four, 
which means 
that any renormalizable QFT of gravity will contain them \cite{tHooft:2011aa}.
The main reason that QG is renormalizable is that it has a built-in
ultraviolet (UV) regulator,
the spin-two ghost field \cite{Stelle:1976gc}, 
which however endangers the unitarity of the theory.
The origin of the good side and the bad side is the same; 
fourth-order derivatives
in the equations of motion. 
 The propagator of a fourth-order theory can be written as a sum
of two propagators of a second-order theory, where
one of them must come with an opposite sign,  where
the wrong-sign propagator  originates from the wrong sign of the kinetic term.
This implies that there exists a negative-metric ghost.
\footnote{That is, 
we have  to deal with  an indefinite metric space 
when the theory is quantized \cite{Boulware:1983td},
instead of the classical Ostrogradsky instability problem
\cite{Ostrogradsky:1850fid,Woodard:2015zca}.}
Our interest in this paper is focused on this troublesome feature, and 
we shall investigate the unitarity problem in  great detail.

Long before Stelle's proof for renormalizability of QG,
Lee and Wick \cite{Lee:1969fy,Lee:1969zze,Lee:1970iw}
used the aforementioned built-in mechanism
to soften the UV behavior of QFT. They  considered
a QED-like theory \cite{Lee:1970iw} with a spin-one massive  ghost field 
and made a very remarkable observation:
The radiative corrections to the ghost propagator due to light fermion (e.g., lepton pair) loops  
makes the single pole split into two complex conjugate poles.
That is, the mass and energy of the ghost become {\it complex}.
This fact became an essential part of their proof of the unitarity:
Ghosts cannot be produced in a scattering process of physical particles 
possessing {\it real} energies  
because of the conservation of energy (in particular, because of the 
conservation of the imaginary part of the energy).
\footnote{This reasoning has been tacitly approved also  by, e.g., 
Coleman \cite{Coleman:1969xz} and  Nakanishi \cite{Nakanishi:1971jj,Gleeson:1971cvx},
 despite the fact that Coleman pointed out
a possible violation of causality in the presence of 
indefinite-metric ghosts,  and that Nakanishi showed violation
of Lorentz invariance in the Lee-Wick prescription of how
to integrate the internal momenta.} If the ghosts are not produced from 
any physical initial states, 
the unitarity of the S-matrix restricted to the physical particles alone, 
i.e., the physical unitarity, holds. 
Although this physical intuition sounds correct,
it seems that Feynman diagrams do not share this intuition since
one has to give a precise prescription 
how to integrate internal momenta 
\cite{Lee:1969fy,Lee:1969zze,Lee:1970iw,Cutkosky:1969fq}.
In fact, 
how  to
choose an appropriate integration contour in the complex energy plane
has been the target of theoretical physicists since then and even recently
\cite{Anselmi:2017ygm,Anselmi:2018kgz,Donoghue:2019fcb,Donoghue:2021eto}.
\footnote{ There are a number of other interesting attempts to show the unitarity,
see for instance \cite{Bender:2007wu,Bender:2007nj,Grinstein:2008bg,Mannheim:2018ljq,Salvio:2019wcp,
Platania:2020knd,Platania:2022gtt}. However, to our knowledge, 
 none of them
are solely based on QFT because certain assumptions have to be made
and it is not clear that they are consistent in QFT, especially in the presence of
interactions such as gravity.
}
The approach described above is based on the standpoint
 that a theory is defined by its Feynman expansion \cite{Coleman:1969xz}.

Our interest in this paper is to investigate the aforementioned 
unitarity problem solely based on QFT.
We will use the tools of QFT, no more, no less.
How to integrate the loop momenta
in a Feynman diagram, apart from its regularization, is dictated by QFT
with no room left for speculation.
If energy is complex, we encounter a mathematical expression,
a complex distribution, which to our knowledge has not been sufficiently explored  
so far aside from a brief discussion in Nakanishi's work
\cite{Nakanishi:1958,Nakanishi:1958-2,Nakanishi:1972wx}.
It is just the integral
\begin{equation} 
\lim_{T\to \infty}\,\frac{1}{2\pi}\int_{-T}^Tdt \,e^{-iE t} =: \delta_\rmc(E)\,,
\label{eq:comp-delta}
\end{equation} 
which appears at each vertex,  where $E$ is the sum $\sum_i E_i$ of the 
energies of particles $i$ entering the vertex. 
If the energy $E$ is real, the limit  $T\to \infty$ is the Dirac delta function,
as we all know. But what is the limit if $E$ is complex? 
Nakanishi \cite{Nakanishi:1958,Nakanishi:1958-2,Nakanishi:1972wx} called it a 
{\it complex delta function} and derived some of its mathematical properties,
but he did not recognize that the complex distribution 
plays an essential role in investigating
 unitarity. The integral (\ref{eq:comp-delta}) with a complex $E$ appears at each vertex of the Dyson expansion in perturbation theory if ghost lines are attached.
Usually, we do not care about this integral because it just gives a 
delta function which expresses the energy conservation at each vertex.
Thus, the mathematical property of the integral (\ref{eq:comp-delta}) is 
intimately related to the energy conservation, especially in the case that
the energy is complex. Or, speaking more strongly, the property of 
the complex delta function (or distribution) defined by the integral 
(\ref{eq:comp-delta}) is all of the content of the ``energy 
conservation law" in the case where complex ghosts participate. 

We will see that $\delta_\rmc(E)$ is non-vanishing even when $\Im E\not=0$, thus 
implying that 
ghosts can be produced with finite (i.e., non-zero) probability through 
a collision of physical particles.  
This leads us to the conclusion that the physical unitarity of the S-matrix
in fourth-order derivative theories
such as the Lee-Wick finite QED and QG is violated in QFT.
\footnote{By the physical unitarity we mean that the probability interpretation
of quantum theory is possible. The physical unitarity of the S-matrix should be distinguished
from the total S-matrix unitarity $(S^\dag S=SS^\dag =1$) which is satisfied 
if the Hamiltonian is
hermitian. In the following discussions we sometimes suppress
``physical'' if it does not lead to a confusion.}
A positive message, however, is that these theories are good effective theories
in  the sense that the physical unitarity of the S-matrix is satisfied below 
a definite threshold  energy for the ghost production, 
where internal ghost propagators in a Feynman diagram  may still be present.
\footnote{Because of the imaginary part of the 
complex mass, the meaning of the threshold should be slightly modified.}

In connection with the property of the complex delta function, we should 
mention a very important point which invalidates the usual 
Feynman rule with energy-momentum conservation used at each vertex  in advance.
For instance, consider a self-energy type one-loop diagram consisting of a 
physical particle $\psi$ with real mass squared $\mu^2$ and a ghost $\varphi$ 
with complex mass squared $M^2$. Many would naively write down the 
expression
\begin{equation}
\Sigma(p) \propto \int d^4q D_\varphi(q)\,D_\psi(p-q)\,  
= \int d^4q \, \frac1{q^2+M^2}\,\frac1{(p-q)^2+\mu^2}\,, 
\label{eq:wrong}
\end{equation}
but this is wrong! T.\ D.\ Lee also started with this expression and correctly 
calculated this integral, even taking the deformed $q^0$-integration 
contour into account properly, 
and reached the conclusion that the amplitude 
has no imaginary part so that the production of a ghost $\varphi$ and 
a physical particle $\psi$ would not occur. This is an incorrect  conclusion 
obtained from a correct calculation but from the wrong starting expression.
The correct expression is not Eq.~(\ref{eq:wrong}), but 
\begin{equation}
\Sigma(p) \propto \int d^4k\,d^4q\, D_\varphi(q)\,D_\psi(k)\, \delta_\rmc(k^0+q^0-p^0)
\delta^3(\bfk+\bfq-\bfp)\,.\label{eq:correct}
\end{equation}
The $k^0$-integration of the complex delta function $\delta_\rmc(k^0+q^0-p^0)$ does 
not give the substitution rule $k^0 \rightarrow p^0-q^0$ alone when the multiplied 
function is not analytic in $k^0$. In the usual Feynman graph case, 
the multiplied function is the propagator, here $D_\psi(k)$, which is not analytic but 
meromorphic in $k^0$. The pole singularities also give additional contributions. 

This problem is serious since the Feynman rules with energy-momentum 
conservation like in Eq.~(\ref{eq:wrong}) are already assumed in all 
the approaches discussing the integration contours in the complex 
energy plane. This implies that those approaches have no ground in QFT. 

We organize the paper as follows. In order to discuss the complex ghost problem properly in QFT, we adopt 
in this paper Lee's purely scalar field theory model \cite{Lee:1969zze}, which Lee devised 
to mimic the essential features of the fourth-order derivative system 
like finite QED or QG. We will present the Lee model in Section 
\ref{sec:LeeModel} and its quantization as an indefinite 
metric QFT in the manner given by Nakanishi \cite{Nakanishi:1972wx}. 
In particular, we will explain the 
unfamiliar metric structure of the complex ghost field $\varphi$ and 
derive two expressions for the ghost propagator; 
the $3d$ and $4d$ momentum expressions. In the latter, as we 
shall see, the integration contour of the zeroth component $k^0$ must take 
a detour around the complex poles that is much deviated from the real axis. This 
causes a considerable complication in the Feynman diagram computations 
despite the apparently covariant compact $4d$-momentum expressions. 

Our main concern is  whether the ghost and anti-ghost are really 
created in 
the scattering processes of physical particles.     
In Section \ref{sec:single-ghost}, we examine the simplest process of single ghost/anti-ghost 
production by the scattering of two physical particles, 
$\psi+\psi \rightarrow \varphi/\varphi^\dagger$.  We calculate this production probability 
to the lowest order
in three ways. First is the direct calculation of the production amplitude, 
given in subsection  \ref{sec:ghost-production}, 
which is almost trivial and simply given by the complex delta function 
$\delta_\rmc$, so we give its precise definition and derive some basic properties
in subsection \ref{sec:complex-delta}. 
We then explicitly show that the ghost is actually produced with 
non-vanishing probability by the two physical particle scattering if the
incident energy $E$ is above $\Re\sqrt{M^2}-\Im\sqrt{M^2}$ (lower threshold) 
and below $\Re\sqrt{M^2}+\Im\sqrt{M^2}$ (upper threshold) with $M^2$ being the complex 
mass squared of the ghost. 

Although this is enough for a proof of ghost appearance, we also calculate 
this production probability in subsection \ref{sec:imaginary-part} by 
computing the imaginary part of the 
forward scattering amplitude of $\psi+\psi \rightarrow \psi+\psi$ with the ghost/anti-ghost
intermediate line (propagator) in the $s$-channel. We present 
two methods of computation for this by using $3d$- and $4d$-momentum expressions 
for the ghost propagator in subsubsections \ref{sec:3d-propagator} and 
\ref{sec:4d-propagator}.
These calculations give the same result as the direct calculation 
in subsection \ref{sec:ghost-production}, as a result of the optical theorem. 
These calculations are presented not only for checking the mutual 
consistency between various ways of calculation, but also for showing how 
the calculations of Feynman graphs should be performed in the presence 
of complex ghost fields and complex delta functions. 

We consider a two ghost production process from a two physical particle 
scattering in Section \ref{sec:pair-production}. The direct calculation of the production 
probability is the simplest, but it is deferred to subsection \ref{sec:direct-production}. 
In the first subsection \ref{sec:forward scattering}, we present the calculation of the forward 
scattering amplitude $\psi+\psi\rightarrow\psi+\psi$ at one-loop in which a pair of 
ghosts/anti-ghosts circulates. We present this to demonstrate how the proper 
calculation goes since this type of loop diagram has been discussed 
by many others, whose calculations 
 often have the problem 
mentioned above in Eq.~(\ref{eq:wrong}) already at the starting expression.
The calculation  using $3d$-momentum expression for the ghost propagator 
is presented in subsubsection \ref{sec:3d-propagator2}, which is simpler compared with the one 
using $4d$-momentum expression presented in \ref{sec:4d-propagator2}. 
The latter calculation 
is actually rather tough and lengthy and its main body has been
moved to the Appendix. 
Instead, we add there a concise explanation for the reason why the naive 
Feynman rule 
(\ref{eq:wrong}) is wrong and how it should be modified, since such 
explanations given for many examples are buried in the lengthy 
calculations moved to the Appendix.

Those three ways of computation give the same 
result as given in subsection \ref{sec:direct-production} for the production probability of a
ghost pair. This result contains the complex delta function which is 
integrated with respect to the 3-momentum $\bfq$ of ghosts. 
Since the complex delta function indirectly depends on $\bfq$ through 
the ghost energy $\omega_\bfq=\sqrt{\bfq^2+M^2}$ as $\delta_\rmc(E-\omega_\bfq)$, 
the $q$-integration reveals a new interesting aspect of the complex 
delta function $\delta_\rmc$.  So, in subsection \ref{sec:explicit-evaluation}, we discuss how to 
evaluate the $q$-integral and obtain the clear result that 
the production probability vanishes for the incident energy 
below the lower threshold and becomes a well-defined finite value 
for energy above the upper threshold. For energy $E$  between the upper 
and lower thresholds, the result is divergent 
though the smearing of the incident energy with any finite width gives 
a well-defined production probability, as is usual for the distribution. 

Section \ref{sec:conclusion} is devoted to the conclusion.

\section{Lee's Model} \label{sec:LeeModel}

In order to examine the properties of complex ghost fields 
in as simple a manner as possible, Lee devised in Ref.\, \cite{Lee:1969zze}
a  purely scalar field model which mimics the essential features of 
the Lee-Wick's finite QED theory in this respect. We call it Lee's model here
and explain this model in a clear manner as Nakanishi presented in Ref.\,
\cite{Nakanishi:1972wx}.

The system consists of three real scalar fields 
$A, B$ and $C$ in the `photon' sector and a normal scalar field $\psi$ in the matter sector, whose free  Lagrangian is given by
\begin{align}
{\cal L}_{\rm free} &= {\cal L}_{\rm ABC} + {\cal L}_{\rm matter}\,, \\ 
{\cal L}_{\rm ABC}&=-\frac12 \left[(\partial_\mu A)^2 + \delta^2A^2\right]
+\frac12 \left[(\partial_\mu B)^2 + m^2B^2\right]
-\frac12 \left[(\partial_\mu C)^2 + m^2C^2\right] -\gamma^2BC \,, \nn
{\cal L}_{\rm matter} &= -\half (\partial_\mu\psi)^2 -\half \mu^2\psi^2\,.
\label{eq:L0}
\end{align}
Here, $A(x)$ is an analogue  of the `photon' $A_\mu$ in QED 
that possesses a small mass $\delta$, 
and $B(x)$ is a negative metric regulator field with a large mass $m$ 
accompanying the `photon' $A$. The $C(x)$ (which   mixes with $B$ and
is absent in the original 
Lee-Wick QED)   is introduced to simulate the continuum 
states like lepton pairs,  $e^+e^-$ and $\mu^+\mu^-$. The mixing 
is represented by the last term $-\gamma^2 BC$. 
We are using the space-favored metric $\eta_{\mu\nu}={\rm diag}(-1, +1,+1,+1)$, 
so that the `photon' $A$ and continuum $C$ fields, as well as the matter field $\psi$, 
are of positive metric  
while the regulator $B$ is of negative metric. We also note that all the fields are real (i.e., hermitian)\footnote{Our 
$B$ field here stands for $i$ times Lee's $B$ field; 
$B_{\rm here}=i B_{\rm Lee}$. We also avoid the use of the extraneous 
(and sometimes confusing) metric operator $\eta$ 
to treat negative metric field $B$. Lee's $\eta^{-1}\phi^\dagger\eta$ is 
simply our $\phi^\dagger$.}, and that the Lagrangian (\ref{eq:L0}) is hermitian 
if the mass parameters $\delta$, $m$ and the $B$-$C$ mixing $\gamma$ are
real.

The $BC$ sector of the free field Lagrangian can also be diagonalized by 
introducing the following complex ghost field $\varphi(x)$:\footnote{%
This complex ghost field $\varphi(x)$ multiplied by $i$ is
Nakanishi's field 
$\phi=(\varphi_1+i\varphi_2)/\sqrt2$  
originally introduced in Ref.\cite{Nakanishi:1972wx}. 
His $\varphi_1$ and $\varphi_2$ are Lee's $C$ and $B$, respectively.}
\begin{eqnarray}
\varphi= \frac1{\sqrt2}(B-iC) 
\qquad \hbox{or} \qquad 
\begin{cases}
B = (\varphi+\varphi^\dagger)/\sqrt2 &\\
C= i (\varphi-\varphi^\dagger) /\sqrt2
\end{cases} \ .
\end{eqnarray}
Then, we have
\begin{eqnarray}
{\cal L}_{BC}
&=& \frac12\left[(\partial_\mu B)^2 - (\partial_\mu C)^2 + m^2(B^2-C^2) - 2\gamma^2 BC \right] \nn
&=& \frac12\left[\partial_\mu\varphi\,\partial^\mu\varphi + M^2\varphi^2  
+\partial_\mu\varphi^\dagger\,\partial^\mu\varphi^\dagger+ {M^*}^2{\varphi^\dagger}^2  \right] ,
\label{eq:LBC}
\end{eqnarray}
where the mass squared $M^2$ for the $\varphi$ field now takes a complex value:
\begin{equation}
M^2 = m^2 + i\gamma^2\,.
\end{equation}
As shown by Nakanishi \cite{Nakanishi:1972wx}, the complex ghost field 
$\varphi$ can  be canonically quantized  and is expanded as 
\begin{equation}
\varphi(x) = \int\frac{d^3\bfq}{\sqrt{(2\pi)^32\omega_{\bfq}}}
\left(
\alpha(\bfq)e^{i\bfq\bfx-i\omega_{\bfq}x^0} 
+ \beta^\dagger(\bfq)e^{-i\bfq\bfx+i\omega_{\bfq}x^0}
\right)\,,
\label{eq:varphi}
\end{equation}
where $\omega_{\bfq}$ is the complex energy 
\begin{equation}
\omega_{\bfq} = \sqrt{ \bfq^2 + M^2 }
= \sqrt{ \bfq^2 + m^2 + i\gamma^2 }\,,
\end{equation}
and the creation and annihilation operators satisfy the off-diagonal 
commutation relations:\footnote{Since $i\varphi_{\rm here}(x)
= \phi_{\rm Nakanishi}(x)$, our creation and annihilation operators 
differ from Nakanishi's by a factor 
of $i$ as $\alpha_{\rm here}(\bfq)= -i\alpha_{\rm Nakanishi}(\bfq)$ and 
$\beta_{\rm here}(\bfq)= i\beta_{\rm Nakanishi}(\bfq)$.}
\begin{eqnarray}
&&[\alpha(\bfp), \beta^\dagger(\bfq)] = [\beta(\bfp), \alpha^\dagger(\bfq)]= -\delta^3(\bfp- \bfq)\,, \nn
&&[\alpha(\bfp), \alpha^\dagger(\bfq)] =[\beta(\bfp), \beta^\dagger(\bfq)]= 0\,.
\label{eq:CCR}
\end{eqnarray}

The Hamiltonian in the BC sector is constructed as usual from the Lagrangian 
(\ref{eq:LBC}) and is given by
\begin{eqnarray}
H_{\rm BC}
&=& \int d^3\bfx \Big(-\frac12\Big)\left[ \pi^2_\varphi + \nabla\varphi\cdot\nabla\varphi+{\rm h.c.}\right] \nn
&=& \int d^3\bfp \left[ -\omega_\bfp \beta^\dagger(\bfp)\alpha(\bfp) - \omega^*_\bfp \alpha^\dagger(\bfp)\beta(\bfp)\right]\ .
\end{eqnarray}
From the commutation relations (\ref{eq:CCR}) and this Hamiltonian, 
we see that 
the 1-particle ghost states
\begin{equation}
\ket{\alpha(\bfp)} := \alpha^\dagger(\bfp)\ket0 \qquad 
\ket{\beta(\bfp)} := \beta^\dagger(\bfp)\ket0
\end{equation}
yield energy eigenstates with eigenvalues $\omega_\bfp^*$ and $\omega_\bfp$ 
respectively, 
\begin{equation}
H_{\rm BC} \ket{\alpha(\bfp)}= \omega_\bfp^* \ket{\alpha(\bfp)}, \qquad 
H_{\rm BC} \ket{\beta(\bfp)}= \omega_\bfp \ket{\beta(\bfp)}\,, 
\end{equation}
and possess the following inner products and norm properties:
\begin{eqnarray}
\langle\alpha(\bfp) \ket{\alpha(\bfq)} &=& 0, \qquad 
\langle\beta(\bfp) \ket{\alpha(\bfq)} = -\delta^3(\bfp-\bfq)\,, \\
\langle\beta(\bfp) \ket{\beta(\bfq)} &=& 0\,, \qquad 
\langle\alpha(\bfp) \ket{\beta(\bfq)} = -\delta^3(\bfp-\bfq)\,.
\end{eqnarray}
That is, these energy eigenstates are themselves zero-norm states and 
have non-vanishing cross-innerproducts between two eigenstates belonging to 
mutually complex conjugate eigenvalues. This is a general property among 
energy eigenstates possessing complex eigenvalues of hermitian Hamiltonians. 
Indeed, for two eigenvectors $\ket{A}$ and $\ket{B}$ 
corresponding to two complex eigenvalues $E_A(\not=E_A^*)$ and $E_B(\not=E_B^*)$ of 
a hermitian Hamiltonian $H$ that satisfy
\begin{equation}
H\ket{A}=E_A\ket{A}\ \rightarrow\ \bra{A} H = \bra{A} E_A^*\,, \qquad 
H\ket{B}=E_B\ket{B}\,, 
\end{equation} 
we have 
\begin{equation}
\bra{A} H \ket{B} = E_A^* \VEV{A|B} = E_B\,\VEV{A|B} \quad \rightarrow\quad 
(E_A^* - E_B)  \VEV{A|B} =0\,.
\end{equation}
This implies that the innerproduct $\VEV{A|B}$ can be non-vanishing only between 
two eigenvectors belonging to mutually complex conjugate eigenvalues. This further implies that any eigenvector of a complex eigenvalue 
necessarily has {\it zero-norm}: $\VEV{E|E}=0$ provided that $E\not=E^*$.

A comment  is in order here on the time evolution of complex ghost 
states. One may think it problematic that the ghost state $\ket{\beta(\bfp)}$ 
evolves as $e^{-i\omega_\bfp t}\ket{\beta(\bfp)}$ with a coefficient 
that blows up exponentially as $t\rightarrow\infty$ since ${\rm Im}(\omega_\bfp)$ is positive. 
The anti-ghost $\ket{\alpha(\bfp)}$, on the other hand, might be thought 
unimportant since it evolves as $e^{-i\omega^*_\bfp t}\ket{\alpha(\bfp)}$ with  
coefficient decreasing exponentially as $t\rightarrow\infty$. However, it is important to
 note that 
such exponential blowing-up or decreasing of the coefficient does not 
imply that the same is true for
the norm of the corresponding state. Indeed, since the ghost $\ket{\beta(\bfp)}$ and 
its ant-ghost $\ket{\alpha(\bfp)}$ have non-vanishing innerproduct only with each 
other, one could define their creation/annihilation operators at time $t$ as
\begin{align}
&\beta^\dagger(\bfp; t) := e^{-i\omega_\bfp t}\beta^\dagger(\bfp)\,,  \qquad 
\beta(\bfp; t) := e^{+i\omega^*_\bfp t}\beta(\bfp), \nn
&\alpha^\dagger(\bfp; t) := e^{-i\omega^*_\bfp t}\alpha^\dagger(\bfp)\,,  \qquad 
\alpha(\bfp; t) := e^{+i\omega_\bfp t}\alpha(\bfp)\,.
\end{align} 
Then, despite the fact that these operators have increasing/decreasing coefficients, 
they produce a superposition state of ghost and anti-ghost, 
$\ket{B(\bfp; t)}$, whose norm is {\it independent} of $t$:
\begin{align}
&\ket{B(\bfp; t)} := \frac1{\sqrt2}(\alpha^\dagger(\bfp; t)+\beta^\dagger(\bfp; t))\ket0 \,,\nn
&\VEV{B(\bfp; t)| B(\bfq; t)} = 
\half\left( \VEV{\alpha(\bfp; t)| \beta(\bfq; t)} 
+ \VEV{\beta(\bfp; t)| \alpha(\bfq; t)}\right) \nn
&=\half\left( \bra{\alpha(\bfp)} e^{+i\omega_\bfp t}\ e^{-i\omega_\bfq t} \ket{\beta(\bfq)} 
+ \bra{\beta(\bfp)}e^{+i\omega^*_\bfp t}\ e^{-i\omega^*_\bfq t}\ket{\alpha(\bfq)}\right) 
 = -\delta^3(\bfp-\bfq) \,,
\end{align}
where we recall that $\ket{B(\bfp; t=0)}$ is just the state created by the 
Schr\"odinger field 
$B=(\varphi+\varphi^\dagger)/\sqrt2$ at $t=0$.
As we will address shortly, it is important that the complex ghost appears only 
in this {\it real} combination (superposition) in the interaction 
Lagrangian. 
The `photon' $A(x)$ and the matter $\psi(x)$ are normal positive metric 
fields which are expanded as usual as follows:
\begin{align}
A(x) &= \int\frac{d^3\bfq}{\sqrt{(2\pi)^32\nu_{\bfq}}}
\left(
a(\bfq)e^{i\bfq\bfx-i\nu_{\bfq}x^0} 
+ a^\dagger(\bfq)e^{-i\bfq\bfx+i\nu_{\bfq}x^0}
\right) , \quad \nu_\bfq= \sqrt{\bfq^2+ \delta^2} \nn
\psi(x) &= \int\frac{d^3\bfp}{\sqrt{(2\pi)^32E_{\bfp}}}
\left(
b(\bfp)e^{i\bfp\bfx-iE_{\bfp}x^0} 
+ b^\dagger(\bfp)e^{-i\bfp\bfx+i\nu_{\bfp}x^0}
\right), \quad E_\bfp= \sqrt{\bfp^2+ \mu^2} \,.
\label{eq:psi}
\end{align}

The Lagrangian of the entire system is assumed to be of the form
\begin{equation}
{\cal L} = {\cal L}_{\rm ABC}+ {\cal L}_{\rm matter}(\psi) + {\cal L}_{\rm int}(\psi, \phi)\,,
\end{equation}
where it is important that the interaction Lagrangian ${\cal L}_{\rm int}$ 
depends on the ABC fields {\it only though    the combined field}  
\begin{equation}
\phi= A + B = A + \frac1{\sqrt2}\left(\varphi+\varphi^\dagger\right).   
\label{phiField}
\end{equation}
That is, the 'photon' $A$ is always accompanied by the complex ghost 
$\varphi$ and the anti-ghost $\varphi^\dagger$,  
so that the complex ghosts are created and annihilated always in a 
{\it real} superposition $\varphi+\varphi^\dagger=\sqrt2 B$. Examples of the interaction Lagrangian which we employ in this paper 
are
\begin{equation}
{\cal L}_{\rm int}(\psi,\phi) = f \psi^3\phi, \ \hbox{or}\ \ f \psi^2\phi^2\,.
\end{equation}
We consider these interactions in perturbation theory by going to the 
interaction picture in which the unitary time evolution operator 
$U(t, t_0)$ is given by
\begin{equation}
U(t, t_0) = {\rm T} \exp\left[ 
i\,\int^t_{t_0} d^4x \, {\cal L}_{\rm int}(\psi(x), \phi(x))\right]. 
\label{U-opr}
\end{equation}
The fields $\psi(x)$ and $\phi(x)$ appearing here are free fields whose explicit 
forms are given above.
The `photon' propagator thus always appears in the form of a
$\phi$ propagator, which is given by
\begin{align}
\bra0 \rmT \phi(x)\,\phi(y) \ket0 
&=
\bra0 \rmT A(x)\,A(y) \ket0
+\half\Bigl(\bra0 \rmT \varphi(x)\,\varphi(y) \ket0 
+\bra0 \rmT \varphi^\dagger(x)\,\varphi^\dagger(y) \ket0 \Bigr) \nn
&=
\int{d^4q\over i(2\pi)^4} e^{iq(x-y)}\left[ \frac1{q^2+\delta^2-i\varepsilon}
-\half\Bigl( \frac1{q^2+M^2} +\frac1{q^2+{M^*}^2} \Bigr) \right] \,.
\label{phiProp}
\end{align}
Note, however, that despite its covariant appearance,  the integration 
contour over $q^0$ is {\it not} along the real axis for the 
ghost propagator part
$\propto1/(q^2+M^2)$, while it is so for the `photon' part 
$\propto1/(q^2+\delta^2-i\varepsilon)$ and the anti-ghost part 
$\propto1/(q^2+{M^*}^2)$. 
\footnote{If we add the second and third term naively, we have a real expression,
which is exactly the fakeon propagator \cite{Anselmi:2018kgz}.}

\begin{figure}[ht]
\begin{center}
\hspace{0.7cm}
\includegraphics[width=7.5cm]{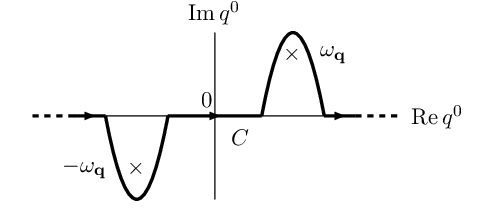}
\includegraphics[width=7.5cm]{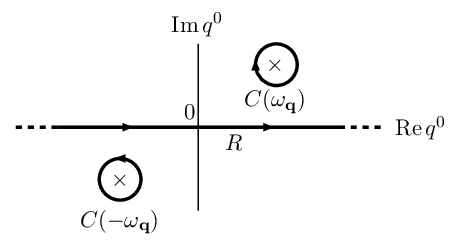}
\caption{Left: The integration contour $C$ for the ghost propagator.
Right: The deformed integration contour corresponding to
(\ref{deform2C2}).}
\label{figC}
\end{center}
\end{figure}

In order to see this, we may calculate, in particular, the propagator of the 
ghost field $\varphi$ explicitly by using the plane wave expansion 
(\ref{eq:varphi}) of $\varphi(x)$. Recalling the definition 
of the T-product and using the commutation relation (\ref{eq:CCR}), we find
\begin{align}
&\bra0 \rmT \varphi(x)\,\varphi(y) \ket0 \nn
&=
\int{d^3\bfq d^3\bfp\over(2\pi)^3\sqrt{2\omega_\bfq2\omega_\bfp}} 
\biggl\{
\theta(x^0-y^0) e^{i(\bfq\bfx-\omega_\bfq x^0)-i(\bfp\bfy-\omega_\bfp y^0)} 
\bra0 \alpha(\bfq) \beta^\dagger(\bfp) \ket0  \nn
&\hspace{10em}{}+ \theta(y^0-x^0) e^{i(\bfp\bfy-\omega_\bfp y^0)
-i(\bfq\bfx-\omega_\bfq x^0)} 
\bra0 \alpha(\bfp) \beta^\dagger(\bfq)\ket0 
\biggr\} \nn
&=
-\int{d^3\bfq\over(2\pi)^32\omega_\bfq} 
\Bigl\{
\theta(x^0-y^0) e^{i\bfq(\bfx-\bfy)-i\omega_\bfq (x^0-y^0)} 
+\theta(y^0-x^0) e^{-i\bfq(\bfx-\bfy)+i\omega_\bfq(x^0-y^0)} 
\Bigr\}\,.
\label{varphiProp}
\end{align} 
Note that the over-all minus sign has come from the negative norm 
commutation relation (\ref{eq:CCR}) of the ghost field.  
In order to rewrite this last expression of 3d integration over $d^3\bfq$ 
into the usual 4d integration form $d^3\bfq dq^0$ by introducing $q^0$ 
variable as
\begin{equation}
= 
-\int{d^3q\over(2\pi)^3} e^{i\bfq(\bfx-\bfy)}\left[ 
\int_C{dq^0\over2\pi i} \frac{e^{-iq^0(x^0-y^0)}}{q^2+M^2}\right]\,, 
\label{GhostProp}
\end{equation} 
the $q^0$-integration contour $C$ must be the one as drawn in Fig.~\ref{figC} (left)
which passes below the 
left pole at $q^0= -\omega_\bfq$ and above the right pole at $q^0=+\omega_\bfq$. 
With this contour we can pick up the right pole giving the 
positive frequency part $e^{-i\omega_\bfq t}$ 
for $t=x^0-y^0>0$, since then the $q^0$ integration contour can be closed 
by adding the half circle at infinity in the lower half plane. 
This is as usual, but, unlike the usual real energy case, 
the energy $\omega_\bfq$ for the ghost $\varphi$ has finite positive imaginary 
part so that the left pole 
is placed much below the real axis and right pole much above the real axis, 
implying $C$  becomes extremely deformed for the ghost case as drawn in 
Fig.~\ref{figC} (left).
This situation becomes opposite for the anti-ghost $\varphi^\dagger$, i.e.,
the left pole at $q^0=-\omega^*_\bfq$ is already placed above, while 
the right pole is placed below the real axis so that the normal integral along the 
real axis satisfies the required property. 
For the 'photon' field $A$, however, both poles are on the real axis
so that we put the usual $-i\varepsilon$ with the mass squared to specify the 
integration contour properly. Thus the proper form of $\phi$-propagator 
is written as 
\begin{align}
&\bra0 \rmT \phi(x)\,\phi(y) \ket0 
=
\sum_{\phi_a=A,\varphi,\varphi^\dagger} \eta_a (-1)^{|a|}\bra{0} \rmT \phi_a(x)\,\phi_a(y) \ket{0} \nn
&\quad{} =
\sum_{\phi_a=A,\varphi,\varphi^\dagger}
\eta_a 
\int{d^3\bfq\over(2\pi)^32\omega^a_\bfq} 
\Bigl\{
\theta(x^0-y^0) e^{i\widehat{q}_a(x-y)} 
+\theta(y^0-x^0) e^{-i\widehat{q}_a(x-y)} 
\Bigr\}
\label{3d-Prop}\\
&\quad {}=\int{d^3\bfq\over i(2\pi)^4} \left[ 
\int_R\frac{dq^0}{q^2+\delta^2-i\varepsilon}
-\frac12\bigg(
\int_C\frac{dq^0}{q^2+M^2} +\int_R\frac{dq^0}{q^2+{M^*}^2}
\bigg) \right] \,e^{iq(x-y)}, 
\label{4d-Prop}
\end{align}
where $\phi_a$ denotes three component fields $(A, \varphi, \varphi^\dagger)$ 
with $\phi=A+(\varphi+\varphi^\dagger)/\sqrt{2}$,
$\eta_a$ is  the weight factor defined as $\eta_a = (+1, -1/2, -1/2)$, 
$(-1)^{|a|}$ is the norm factor $(+1, -1, -1)$
and $\calE^a_\bfq$ and $\what q_a^\mu$ are the energy and the on-shell 
4-momentum defined as 
\begin{equation}
\calE^a_\bfq = \sqrt{\bfq^2+m_a^2}= ( \nu_\bfq,\ \omega_\bfq,\ \omega^*_\bfq)\,, \ \quad 
m_a =(\delta,\ M,\ M^*)\,, \qquad \ 
\what q^\mu_a = (\calE^a_\bfq,\ \bfq\, ) \, .
\label{EnergyOnShellMomentum}
\end{equation}
We use the expressions (\ref{3d-Prop}) and (\ref{4d-Prop}) for 
the $\phi$ propagator in 
our calculations in the following sections. 
The second line of (\ref{3d-Prop}) gives the 3d momentum form
of the $\phi$-propagator with 
on-shell $\widehat{q}_a^0 = \omega_\bfq^a$, which is free from the 
complications of the integration contour of $q^0$, despite the fact that it appears
non-covariant. The third line (\ref{4d-Prop}) gives the 4d momentum form
for which the $q^0$ integration contour should be $R$, $C$ and $R$ when 
$\phi_a$ represents the 'photon' $A$, ghost $\varphi$ and anti-ghost $\varphi^\dagger$, respectively.

For general Feynman diagrams possessing many propagators, the integration
contour for the energy variable $q^0$ for each propagator must satisfy 
such a requirement. When the energy conservation conditions are  
imposed via the vertices, those energy variables become  
dependent on each other, which transforms the requirements on the contours of 
the original energy variables into much more involved conditions of the independent 
energy variables. 
One can avoid such a complicated consideration of the integration contours 
if the 4-th component $q^0$ is not introduced and one uses 
only the 3d momentum variables  in a similar manner to 
the old-fashioned  perturbation theory. 
This method lacks manifest covariance, but the 
deformation of the integration contour of only the energy variables
violates manifest covariance anyway  even  if  the 4d momentum 
expression of propagators is used.

\section{Single ghost production}\label{sec:single-ghost}

Let us now consider the simplest process of single ghost production by 
the collision of two matter particles: $\psi+\psi
\to \phi$. We take an interaction Lagrangian of 
the form
\begin{equation}
{\cal L}_{\rm int} = \frac{f}2 \psi(x)\psi(x)\phi(x) \,,
\label{YukawaInt}
\end{equation}
and calculate the ghost production probability in two ways in this section.

\subsection{Direct calculation of $\phi$-production}\label{sec:ghost-production}

The initial state is taken to be the two particle matter state
\begin{align}
\ket{ I (\bfp_1, \bfp_2) } 
=
\sqrt{(2\pi)^3 2E_{\bfp_1}} b^\dagger(\bfp_1)
\sqrt{(2\pi)^3 2E_{\bfp_2}} b^\dagger(\bfp_2) \ket0 \,.
\label{Init2psi}
\end{align}
To first order in the coupling strength $f$ of Dyson's $S$-matrix, 
$S=U(\infty, -\infty)$, with $U$ being the operator Eq.~(\ref{U-opr}), this initial state evolves
into a single $\phi$-particle state as
\begin{align}
(S-1)^{(1)} \ket{ I (\bfp_1, \bfp_2) } 
&= i \frac{f}2 \int d^4x C(x^0) \psi(x)\psi(x)\phi(x) \ket{ I (\bfp_1, \bfp_2) } \nn
&= i f 
\int dx^0d^3\bfx\, C(x^0) e^{-i(p_1^0+p_2^0 - \widehat{q}^0)x^0} \nn
&\hspace{2em}\times 
\int{d^3\bfq\over\sqrt{(2\pi)^3}}
e^{i(\bfp_1+\bfp_2-\bfq)\bfx} 
\frac1{\sqrt{2\widehat{q}^0}}  
\phi^\dagger(\bfq)\ket0 \quad  + \Bigl( \cdots\Bigr)\,,
\label{eq:Tamp1}
\end{align}
where $p_i^0 = \sqrt{\bfp_i^2 + \mu^2} = E_{\bfp_i}\ (i=1,2) $, and 
$( \cdots)$ represents other states consisting of $(\psi^\dagger)^4\phi^\dagger$ and 
$(\psi^\dagger)^2\phi^\dagger$.  
Here, since the field $\phi$ stands for the linear combination of three fields, 
$A+(\varphi+\varphi^\dagger)/\sqrt2$ in Eq.~(\ref{phiField}), the $\phi^\dagger(\bfq)\ket0$ 
together with the related coefficient is understood to represent the following 
three terms:
\begin{align}
e^{i\widehat{q}^0x^0}\frac1{\sqrt{2\widehat{q}^0}}\phi^\dagger(\bfq)\ket0 
&=
e^{i\nu_\bfq x^0}\frac1{\sqrt{2 \nu_\bfq }}a^\dagger(\bfq)\ket0 \nn
&\qquad{}+ \frac1{\sqrt2} \Big(
e^{i\omega_\bfq x^0}\frac1{\sqrt{2 \omega_\bfq }}\beta^\dagger(\bfq)\ket0 
+ e^{i\omega_\bfq^* x^0}\frac1{\sqrt{2 \omega_\bfq^* }}\alpha^\dagger(\bfq)\ket0 
\Big)\,.
\label{phi-state}
\end{align}
The hat $\what{\phantom{q}}$ attached to $q^0$ 
is a reminder of the fact that it 
changes meaning depending on the following states; i.e., 
$\widehat{q}^0=(\nu_\bfq,\, \omega_\bfq,\, \omega^*_\bfq)$ for the `photon' state
$a^\dagger(\bfq)\ket0$, the ghost $\beta^\dagger(\bfq)\ket0$ and anti-ghost 
$\alpha^\dagger(\bfq)\ket0$, respectively.    

$C(x^0)$ is the regularization factor that avoids the exponential divergence 
difficulty for $x^0 \rightarrow\pm\infty$, as was noted by Lee and Wick. For $C(x^0)$ we can 
take, for example, a sharp cut-off that represents an interaction acting only 
in a finite time interval $x^0\in[-T,\ +T]$:
\begin{equation}
C_T(x^0) =  \theta(x^0+T)\,\theta(T-x^0) \,.
\label{Tcutoff}
\end{equation}
To simplify our analytic treatment, however, we  adopt the following 
adiabatic cut-off which corresponds to a smoother counterpart 
of (\ref{Tcutoff}) with $2T \sim\sqrt{\pi}/a$:
\begin{eqnarray}
C_a(x^0) = e^{-a^2 x_0^2} \,.
\label{AdiabaticCutoff}
\end{eqnarray}
We multiply ${\cal L}_{\rm int}$ in the time evolution $U$ operator (\ref{U-opr}) 
by this 
regularization factor and eventually take the limit $a \rightarrow0$ (or $T\rightarrow\infty$) 
to remove the regularization. Nevertheless, it is important that the 
unitarity of the $U$ operator always holds for any finite $a$ since 
$C(x^0){\cal L}_{\rm int}(x^0)$ is hermitian at any time.    

Performing $d^3\bfx$ integration in Eq.~(\ref{eq:Tamp1}) 
and then $d^3\bfq$ integration, we obtain
\begin{align}
(S-1)^{(1)} \ket{I} 
&= i f 
(2\pi)\Delta_a(p_1^0+p_2^0 - \widehat{q}^0) 
\sqrt{\frac{(2\pi)^3}{2\widehat{q}^0}} 
\phi^\dagger(\bfq)\ket0\Big|_{\bfq=\bfp_1+\bfp_2} \quad  + \Bigl( \cdots\Bigr)\,,
\label{Produced-phi}
\end{align}
where we have introduced a regularized complex delta function
\begin{equation}
\Delta_a(z) 
:= \frac1{2\pi}\int_{-\infty}^\infty dt\, e^{-a^2 t^2} e^{-iz t}  
=\frac1{2\sqrt{\pi}\,a}\,e^{-z^2/4a^2}\,.
\end{equation}
As the adiabatic factor is removed in the  $a\rightarrow0$ limit, this goes to 
the {\it complex delta function} $\delta_{\rm c}(z)$ with 
complex argument $z$, 
\begin{equation}
\lim_{a\rightarrow0}\Delta_a(z) =  \frac1{2\pi}\int_{-\infty}^\infty dt\, e^{-iz t}
 =:  \delta_{\rm c}(z)\,,
\label{Delta-a}
\end{equation}      
which in turn reduces to the usual Dirac's delta $\delta(x)$ for real argument $z=x$, and 
was first introduced by Nakanishi \cite{Nakanishi:1958,Nakanishi:1958-2} long ago
in connection with the construction of Hamiltonian eigenstates of unstable particles 
with complex eigenvalue. We shall show in the next subsection 
 that $\delta_{\rm c}(z)$ 
gives a well-defined {\it distribution} and write henceforth only the 
expressions for the limit $a=0$ 
unless the explicit regularization is necessary.
\begin{figure}[ht]
\begin{center}
\hspace{0.7cm}
\includegraphics[width=6.5cm]{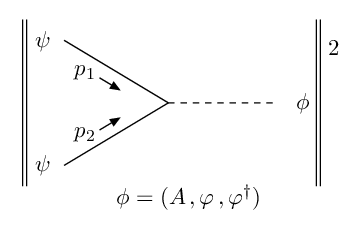}
\includegraphics[width=7.7cm]{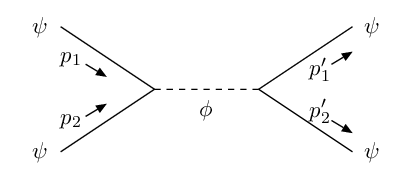}
\caption{Left:  Graphical presentation of the norm
(\ref{phiNorm}). Right: The forward scattering amplitude of two physical particles.}
\label{figN}
\end{center}
\end{figure}

Noting Eq.~(\ref{phi-state}),
using the commutation relation (\ref{eq:CCR}) 
as well as $[\, a(\bfp)\,,\,a^\dag(\bfq)\,]=
\delta^3(\bfp-\bfq)  $ and writing $p_1^\mu+p_2^\mu=P^\mu$, 
we can calculate the norm of the produced $\phi$-state, 
$(S-1)^{(1)}\ket{I}_\phi$ in Eq.~(\ref{Produced-phi}), 
as
\begin{align}
&\big|\!\big|(S-1)^{(1)}\ket{I}_\phi\big|\!\big|^2 
= f^2 
(2\pi)^3\delta^3(\bfP'-\bfP)(2\pi)\delta(P^0-P^0) \nn
&\qquad  {}
\times\bigg\{
2\pi\delta(P^0 - \nu_\bfP) \frac1{2\nu_\bfP}
- \half \Big(
2\pi\delta_{\rm c}(P^0 - \omega_\bfP)\frac1{2\omega_\bfP}
+2\pi\delta_{\rm c}(P^0 - \omega^*_\bfP)\frac1{2\omega^*_\bfP}
\Big)\bigg\}\,,
\label{phiNorm}
\end{align}
where we have distinguished the momentum $\bfP'$ of the bra-state $\bra{\bfP'}$ 
from that $\bfP$ of the ket-state $\ket{\bfP}$ for clarity although 
$\bfP'=\bfP$ when computing the norm $\big|\!\big| \ket{\bfP}\big|\!\big|^2 = \langle\bfP\ket{\bfP}$. 
Fig.~\ref{figN} (left) is a graphical presentation of the norm
(\ref{phiNorm}). We have also used the 
identity
\begin{equation}
\bigl[2\pi\delta_{\rm c}( P^0 - \widehat{q}^0 )\bigr]^2 = 2\pi\delta( P^0 - P^0)\cdot 
2\pi\delta_{\rm c}( P^0 - \widehat{q}^0 ),
\label{Delta0}
\end{equation}
which is shown to hold for the complex delta function 
$\delta_{\rm c}(P^0-\widehat{q}^0)$ with real $P^0$ and complex $\widehat{q}^0$, 
shortly below. 

In the norm expression (\ref{phiNorm}), the factor 
$(2\pi)^3\delta^3(\bfP'-\bfP)(2\pi)\delta(P^0-P^0)=(2\pi)^4\delta^4(0)$ 
is the total space-time volume so that it is divided out for the production 
probability per unit space-time volume, as usual. 
The first term is the norm of the `photon' particle $A$ and 
the second and third terms are that of the ghost and anti-ghost, respectively,
 with negative sign.
The first term 
yields the production probability $P_A$ of `photon' $A$ per unit space-time volume as
\begin{equation}
P_A = f^2 2\pi\delta(P^0 - \nu_\bfP) \frac1{2\nu_\bfP}\,.
\end{equation} 
This implies that the `photon' production occurs only exactly 
at the resonating energy 
$P^0=\nu_\bfP=\sqrt{\bfP^2+\delta^2}$ while the production probability is zero below and 
beyond the threshold $P^0=\nu_\bfP$.  
This term merely reproduces the usual result, 
confirming the validity of this computation. 

The second and third terms, therefore, give the production probability 
$P_\varphi$ of 
our ghost (superposition) $\varphi+\varphi^\dagger$ as 
\begin{equation}
P_\varphi 
=- {\rm Re}\left[f^2 2\pi\delta_\rmc(P^0 - \omega_\bfP) \frac1{2\omega_\bfP}\right]\,.
\label{Pghost}
\end{equation}
We note that this probability is {\it real and negative}
as shown more explicitly in the next subsection, 
thus implying 
a violation of physical unitarity: That is, if it is non-vanishing, 
the probability of other final physical states 
(consisting of physical particles alone) exceeds one.

Before ending this subsection, we add an important remark here:
The opposite process to the single ghost production is
the ``decay'' of the ghost into two physical particles, which of course 
occurs due to time reversal invariance. One may then naturally wonder, are the 
complex ghost particles stable? 
The answer is yes
(in contrast to the assumption of \cite{Donoghue:2019fcb,Donoghue:2021eto}). 
First of all, 
the ghost state created in the superposition $\varphi+\varphi^\dagger$ has 
a negative norm. Since the Dyson's S-matrix of the present system is unitary, 
the negative norm, say $-1$, of the initial ghost state must be conserved. 
So, whatever final states are produced from the initial ghost state, 
the norms of all those final states sum up to the value $-1$ of the initial 
ghost state's norm. To realize this negative value, however, ghost particles 
must be contained among the final states. This implies that the ghosts 
can never disappear by totally `decaying out' into lower mass physical 
particles.

\subsection{Complex delta function}\label{sec:complex-delta}
To see that $P_\varphi $  is actually non-vanishing, 
let us now analyze the property of 
the complex delta function $\delta_{\rm c}(P^0-\omega_\bfP)$ in more detail.  
From the definition (\ref{Delta-a}), we have an expression 
\begin{equation}
\delta_{\rm c}(P^0-\omega_\bfP) = \lim_{a\rightarrow0}\,\Delta_a(P^0- \omega_\bfP)= 
\lim_{a\rightarrow0}\,\frac{1}{2a\sqrt{\pi}}\, 
\exp \Bigl[ - \frac{(P^0-\omega_\bfP)^2}{4a^2}\Bigr]\,,
\end{equation}
where the ghost energy $\omega_\bfP$ is the complex quantity
\begin{equation}
\omega_\bfP = \sqrt{ \bfP^2 + m^2 + i\gamma^2 }\,.
\end{equation}
In the center of mass (CM) frame, $\bfP={\bf 0}$, this $\omega_\bfP$ becomes 
\begin{align}
&\omega\equiv\omega_{\bf 0}= \sqrt{ m^2 + i\gamma^2 } = ( m^4+\gamma^4 )^{1/4} e^{i\vartheta/2}
\quad  
\hbox{with} \quad 
\tan\vartheta= \frac{\gamma^2}{m^2} \nn
&\rightarrow\ \Re \omega=  ( m^4+\gamma^4  )^{1/4}\cos(\vartheta/2)\,, \ \ \ 
\Im \omega=  ( m^4+\gamma^4 )^{1/4}\sin(\vartheta/2)\, .
\end{align}
The $\Delta_a$ function then becomes (denoting $P^0$ simply by $E$) 
\begin{align}
&\Delta_a(E-\omega) = 
\frac1{2a\sqrt{\pi}} \exp\left[-\frac{(E-\Re\omega)^2-(\Im\omega)^2}{4a^2}\right]
\cdot e^{-i\Theta} \label{DeltaAsqr}\\
& \hbox{with} \quad 
\Theta=\frac2{-4a^2}(E-\Re\omega)\Im\omega\,. 
\end{align}
This function, as a function of (real) energy $E$, is clearly non-vanishing 
everywhere for finite $a$, so that the ghost production probability $P_\varphi$ 
in Eq.~(\ref{Pghost}) is 
non-vanishing for any initial energy $E=P^0$ on any finite time interval 
$T \sim a^{-1}$. This may not sound surprising since the energy is not conserved for 
finite time interval in any case.

Now consider the infinite time limit $a\rightarrow0$. Note that the parabolic function 
of $E$ in the exponent in  Eq.~(\ref{DeltaAsqr})
\begin{equation}
(E-\Re\omega)^2-(\Im\omega)^2
=(E-\Re\omega+\Im\omega)(E-\Re\omega-\Im\omega)
\end{equation}
takes negative values only in the interval $\Re\omega-\Im\omega< E <\Re\omega+\Im\omega$ 
and positive values or zero otherwise. Since it is multiplied by the 
factor $-1/a^2$ and hence approaches $-\infty$ as $a\rightarrow0$, 
the $\Delta_a(E-\omega)$ function vanishes  in the infinite time 
limit $a\rightarrow0$
{\it outside}  the finite energy interval of width $2\,\Im\omega$ 
around $E=\Re\omega$, i.e., \ $|E-\Re\omega|<\Im\omega$:
\begin{equation}
\delta_{\rm c}(E-\omega)= \lim_{a\rightarrow0} \Delta_a(E-\omega) = 0 \quad \hbox{for} \quad 
\begin{cases}
E < \Re\omega-\Im\omega\ & \\    
\Re\omega+ \Im\omega< E  & 
\end{cases}
\label{LowerUpperThreshold}
\end{equation}
Inside this interval, on the other hand, 
the function $\Delta_a(E-\omega)$ is 
non-vanishing but diverges while rapidly oscillating as $a\rightarrow0$, 
implying that the limiting function $\delta_{\rm c}(E-\omega)$ is not a function, but rather 
a {\it distribution}. 
To see that it gives a well-defined distribution 
possessing non-vanishing support in the interval $|E-\Re\omega|<\Im\omega$, let us 
compute an integral of it multiplied by a test function $f(E)$ which is 
analytic in a necessary domain:
\begin{equation}
\int_{-\infty}^\infty dE\, \delta_{\rm c}(E-\omega)\, f(E)
=\lim_{a\rightarrow0}\int_{-\infty}^\infty dE \,
\frac1{2a\sqrt{\pi}}\exp\left[-\frac{(E-\omega)^2}{4a^2}\right] f(E) \,.
\end{equation}
\begin{figure}[h]
\begin{center}
\hspace{0.7cm}
\includegraphics[width=7.5cm]{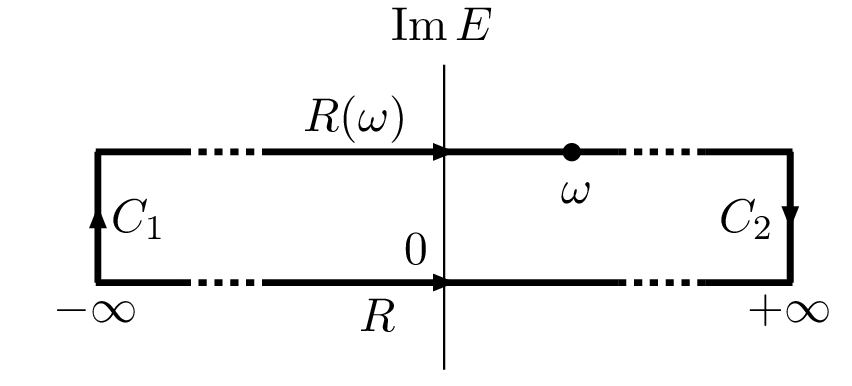}
\caption{The integration contour  (\ref{ContourRomega}). }
\label{contour-R}
\end{center}
\end{figure}
We can deform the contour of this $E$ integral along the real axis $R$ to 
the following:
\begin{align}
R[-\infty\rightarrow+\infty] \ \Rightarrow\ \phantom{+} & 
C_1[ -\infty+i 0 \rightarrow-\infty+i \Im\omega] \nn
{}+{} &R(\omega)[ -\infty+i \Im\omega\rightarrow+\infty+ i \Im\omega] \nn
{}+{} &C_2[ +\infty+i \Im\omega\rightarrow+\infty+ i0 ]\,,
\label{ContourRomega}
\end{align}
where the deformed contour is shown in Fig.~\ref{contour-R}.
The integrals along the finite vertical segments $C_1$ and $C_2$ vanish since 
$\exp[-(E-\omega)^2/a^2]$ vanishes for $\Re E \rightarrow\pm\infty$ 
with  $\Im E$ kept finite. 
The integral can thus be evaluated by the contour integral along $R(\omega)$, which 
denotes the horizontal contour parallel to the real axis $R$ and passing the 
point $z=\omega$ i.e., \ $z = x + i \Im\omega$\ (with real $x\in[-\infty, \infty]$). 
This deformation is allowed when the test function $f(E)$ is 
analytic in the 
complex plane inside the rectangular region surrounded by $R+C_1+R(\omega)+C_2$
(see Fig.~\ref{contour-R}).
The integral along $R(\omega)$ is evaluated by making the change of variable 
$E= E'+ i\Im\omega$ where $E'$ is real on $R(\omega)$ and 
$\delta_{\rm c}(E-\omega)$ reduces to the usual Dirac's delta $\delta(E'- \Re\omega)$ 
since $E-\omega=E'- \Re\omega$ is real on $R(\omega)$.
With this we find
\begin{equation}
\int_{R(\omega)}dE\,\delta_{\rm c}(E-\omega)f(E) = \int_{-\infty}^\infty dE'\, 
\delta(E'-\Re\omega)f(E'+ i \Im\omega)
= f(\Re\omega+i \Im\omega) = f(\omega)\,,
\label{DeltaProperty}
\end{equation}
which proves that the distribution $\delta_{\rm c}(E-\omega)$ works as if it is the 
usual Dirac's delta function despite the fact that $E$ is real and $\omega$ is 
complex. This also proves 
\begin{equation}
\delta_{\rm c}(E-\omega)f(E) = \delta_{\rm c}(E-\omega)f(\omega),
\label{DeltaProperty2}
\end{equation}
which proves Eq.~(\ref{Delta0}) when $f(E)=\delta_{\rm c}(E-\omega)$, 
as promised.  

We should note that the remarkable property (\ref{DeltaProperty}) for 
the complex delta function holds if and only if the test function $f(E)$ 
has no singularity inside the rectangular region stated above
\cite{Nakanishi:1972wx}. In practice 
however, $f(E)$ is usually given by the products of meromorphic Feynman 
propagators which often have pole singularities in the rectangular regions 
in question, in which case one has to pick up that 
contribution also. As we will see soon, this is actually the point touching 
the core of the problem.

In the actual scattering experiment, the total energy $E$ of the initial 
particles, $P^0=p_1^0+p_2^0$, must necessarily have a certain uncertainty 
around  $P^0$, 
which may be described, for instance, by a Gaussian distribution 
with standard deviation 
$\sigma$ of the form
\begin{equation}
f_{P^0}(E) = \frac1{\sqrt{2\pi\sigma^2}}\exp
\left[-\frac12\Bigl(\frac{E-P^0}{\sigma}\Bigr)^2\right] \,.
\label{GaussianSmear}
\end{equation} 
For such initial 
particles, the ghost production probability (\ref{Pghost}) should be averaged 
with this distribution. The actual production rate of complex ghost per unit 
space-time volume is then found to be
\begin{align}
P_\varphi &= -\Re \Bigl[f^22\pi\int_{-\infty}^\infty dE\, f_{P^0}(E) \delta_{\rm c}(E-\omega) \frac1{2\omega}\Bigr] \nn
&=-\Re \Bigl[f^2\frac{\pi}{\omega}f_{P^0}(\omega)\Bigr]
=-\frac{f^2}{\sqrt{2\pi\sigma^2}}\Re\biggl[\frac{\pi}{\omega}\exp
\Bigl[-\frac12\Bigl(\frac{\omega-P^0}{\sigma}\Bigr)^2\Bigr]\biggr]\,,
\label{ghost-production}
\end{align}
which is finite as far as $\sigma$ is finite.

\subsection{Imaginary part calculation of the forward scattering amplitude}
\label{sec:imaginary-part}

Since the interaction Hamiltonian is hermitian, the total $S$-matrix unitarity 
follows trivially and leads to the optical theorem for the $T$-matrix:
\begin{align}
&S = 1 + (2\pi)^4\delta^4(P_F-P_I)\,i\hat{T}\,, \nn
&S^\dagger S= 1 \rightarrow2\Im \bra{I}\hat{T}\ket{I} = (2\pi)^4
 \sum_{F} \delta^4(P_F-P_I)\,\big|\!\big|\,\hat{T}\ket{I}_F \big|\!\big|^2
 \label{OpticalTh}\,,
\end{align}
where $\hat{T}\ket{I}_F$ denotes all independent states $\ket{F}$ 
contained in the final scattered state $\hat{T}\ket{I}$.
In perturbation theory, this optical theorem holds for each set of the 
same order terms in the coupling constant $f$. Additionally, 
the imaginary part of the particular Feynman diagram on the LHS equals the 
RHS with the sum over $F$ restricted to a certain subset. 

Let us now calculate the forward scattering amplitude of two physical particles
 with the initial state $\ket{I(\bfp_1, \bfp_2)}$ in Eq.~(\ref{Init2psi}). At
second order in $f$ of the interaction Lagrangian (\ref{YukawaInt}), 
there are only three diagrams which have $\phi$-propagator in $s$-, $t$- 
and $u$-channels, and we calculate only the $s$-channel amplitude 
since it has an imaginary part that reproduces the $\phi$-production probability in the 
previous subsection \ref{sec:ghost-production}.  Here we note that 
the result Eq.~(\ref{phiNorm}) which we have found as the norm of 
the produced $\phi$-particle state (\ref{eq:Tamp1}) 
in the subsection \ref{sec:ghost-production} can be rewritten in the form of 
the RHS of optical theorem (\ref{OpticalTh}) as
\begin{align}
&(2\pi)^4\sum_{\bfq} \delta^4(P_F-P_I) \big|\!\big| \,\hat{T}\ket{I(\bfp_1,\bfp_2)}_{\text{$\phi(\bfq)$-prod.}} \big|\!\big|^2 \nn
&= 
f^2\,\int\frac{d^3\bfq}{(2\pi)^3} (2\pi)^3\delta^3(\bfq-\bfP) 
\bigg\{
2\pi\delta(\nu_\bfq-P^0) \frac1{2\nu_\bfq}
- \Re \Big[2\pi\delta_{\rm c}(\omega_\bfq-P^0)\frac1{2\omega_\bfq}\Big]
\bigg\}\,.
\label{phiNorm2}
\end{align}
We should find the same result by calculating the imaginary part of 
the forward scattering amplitude on the LHS of the optical theorem 
(\ref{OpticalTh}).
%
The $s$-channel tree level amplitude at $O(f^2)$ corresponding to the 
Feynman diagram Fig.~\ref{figN} (right) is now also easily computed:
\begin{align}
&\bra{I'(\bfp'_1, \bfp'_2)}(S - 1)
\ket{I(\bfp_1, \bfp_2)}
\big|_{\text{$s$-ch.}}
\Bigl( {}=i(2\pi)^4\delta^4(P'-P)\bra{I'(\bfp'_1, \bfp'_2)}\hat{T}\ket{I(\bfp_1, \bfp_2)}
\big|_{\text{$s$-ch.}} \Bigr)\nn
&\quad {}=\bra{I'(\bfp'_1, \bfp'_2)} \frac12 \Big(\frac{if}2\Big)^2 
\rmT \int d^4x\,d^4y\,C_a(x^0)\,\psi^2(x)\phi(x)\,C_a(y^0)\psi^2(y)\phi(y)\ket{I(\bfp_1, \bfp_2)}
\big|_{\text{$s$-ch.}} \nn
&\quad {}=(if)^2 \int d^4x\,d^4y\,e^{-iP'x}C_a(x^0) 
\bra0 \rmT \,\phi(x)\phi(y)\ket0 \,C_a(y^0) e^{iPy}\,,
\label{eq:S-1-0}
\end{align}
where $P^\mu:=p_1^\mu+p_2^\mu, P'^\mu:=p'^\mu_1+p'^\mu_2$ and $C_a(x)$ is the 
adiabatic Gaussian cutoff (\ref{AdiabaticCutoff}). 
We will compute this $S$-matrix element (\ref{eq:S-1-0}) in two ways;
first using the 3d-momentum form (\ref{3d-Prop}) of the $\phi$ propagator, 
and second using 
the covariant 4d-momentum form  (\ref{4d-Prop}).
We will explicitly see why 
a careful treatment of a regularization (cutoff)  (\ref{AdiabaticCutoff}) 
is indispensable for obtaining a coincident final result in both calculations.
Once the equivalence of these methods is established, we can use 
either form of the propagator freely. 


\subsubsection{Use of 3d-momentum form of  the $\phi$ propagator}
\label{sec:3d-propagator}

We first present the calculation using the 3d-momentum form
(\ref{3d-Prop}) of the $\phi$ propagator.
Now, we insert the 3d-momentum form (\ref{3d-Prop}) into 
the propagator $\VEV{0|\rmT \phi(x)\phi(y)|0}$ in Eq.~(\ref{eq:S-1-0}). Then, 
making a change of time 
variables $(x^0, y^0)$ into CM and relative time $(T, t)$
\begin{equation}
\begin{cases}
T= \frac12(x^0+y^0)  \\ 
t= x^0-y^0 
\end{cases}
\rightarrow 
\begin{cases}
x^0 = T + t/2   \\ 
y^0 = T - t/2
\end{cases}\,, \quad  dx^0dy^0= dT dt
\label{CM-time}
\end{equation}
and writing explicitly the adiabatic (Gaussian) cutoff factor 
(\ref{AdiabaticCutoff}) 
at both interaction points $x^0$ and $y^0$,
\begin{equation}
C_a(x^0)C_a(y^0) = 
\exp (-a^2({x^0}^2+{y^0}^2)) = 
\exp \left(-2a^2 T^2 - \frac12 a^2 t^2  \right)\, , 
\end{equation}
we have, after performing the $d^3\bfx d^3\bfy$ integrations and 
using 
$\delta^3(\bfP'-\bfq)\delta^3(\bfq-\bfP)
= \delta^3(\bfP'-\bfP)\delta^3(\bfq-\bfP)$,   
\begin{align}
&\hbox{LHS of (\ref{eq:S-1-0})} \nn
&\hspace{.5em}=
(if)^2\sum_a \eta_a (2\pi)^3\delta^3(\bfP'-\bfP) 
\int\frac{d^3\bfq}{(2\pi)^32\calE_\bfq^a} (2\pi)^3\delta^3(\bfq-\bfP) \nn
&\hspace{1em}\times\int dT \,e^{-2a^2 T^2}e^{i(P'^0-P^0)T}  
\int dt \,e^{-a^2 t^2/2} 
\Bigl(
\theta(t)e^{i(\overline P^0-\calE_\bfq^a)t} +
\theta(-t)e^{i(\overline P^0+\calE_\bfq^a)t}
\Bigr) 
\label{LHSof3.29}
\end{align}
with $\overline P^0 := (P'^0+P^0)/2$. 
Note that the integrations of the CM time $T$ and relative time $t$ are 
totally separated including their adiabatic cutoff factors. 
The $dT$ integration yields the factor
\begin{equation}
\int dT \,e^{-2a^2 T^2}e^{i(P'^0-P^0)T}= 2\pi\Delta_{\sqrt2 a}(P'^0-P^0)
\ \underset {a\rightarrow0}{\longrightarrow }\ 
(2\pi)\delta(P'^0-P^0)\,,
\end{equation}
which together with the above $(2\pi)^3\delta^3(\bfP'-\bfP)$ completes 
the total energy-momentum conservation factor 
$i(2\pi)^4\delta^4(P'-P)$, which should be  divided out when giving a $T$-matrix 
element.  

Next, we consider the $dt$ integration part
(denoting $\bar{P}^0$ by $E$ as in (\ref{DeltaAsqr}))
\begin{eqnarray}
\int dt \,e^{-a^2 t^2/2}
 \Big(\theta(t)e^{i(E-\omega^a_\bfq)t} +\theta(-t)e^{i(E+\omega_\bfq^a)t}\Big) \,.
\label{Thetat-EXPizt}
\end{eqnarray}
For the anti-ghost case first,  $\omega_\bfq^a=\omega^*_\bfq$ 
has negative imaginary part $\Im\omega_\bfq^* <0$. We note that the integral 
converges even in the limit $a\rightarrow0$ since $e^{i(E \mp \omega^*_\bfq)t}$ 
decreases exponentially 
as $t\rightarrow\pm\infty$, meaning that the $a\rightarrow0$ limit is easily calculated 
by setting $a=0$ from the beginning:
\begin{align}
\int dt \, \Big(\theta(t)e^{i(E-\omega^*_\bfq)t} +\theta(-t)e^{i(E+\omega^*_\bfq)t}\Big) 
&= \left[ \frac{e^{i(E-\omega^*_\bfq)t}}{i(E-\omega^*_\bfq)} \right]^\infty_0 
+ \left[ \frac{e^{i(E+\omega_\bfq^*)t}}{i(E+\omega_\bfq^*)} \right] ^0_{-\infty} \nn
&= i\left(\frac1{E-\omega^*_\bfq} -  \frac1{E+\omega_\bfq^*}\right) \,.
\label{ImOmega<0}
\end{align}

In the ghost case, $\omega_\bfq^a=\omega_\bfq$ has {\it positive} 
imaginary part $\Im\omega_\bfq >0$ oppositely to the anti-ghost case. 
So the $a\rightarrow0$ limit is divergent since 
$e^{i(E\mp \omega_\bfq)t}$ exponentially blows up as $t\rightarrow\pm\infty$, respectively.
But this also implies that the functions $e^{i(E\mp \omega_\bfq)t}$ 
decrease exponentially in the opposite directions $t\rightarrow\mp \infty$. 
If we rewrite $\theta(\pm t)$ as  $1 -\theta(\mp t)$  in Eq.~(\ref{Thetat-EXPizt}), 
then those terms with a $-\theta(\mp t)$ part give convergent quantities 
in the $a\rightarrow0$ limit as in the anti-ghost case, while the terms with $1$ 
give the regularized complex delta function; that is, we can evaluate 
Eq.~(\ref{Thetat-EXPizt}) in the ghost case as
\begin{align}
&\lim_{a\rightarrow0}\int dt \, e^{-a^2t^2/2}\Big(\theta(t)e^{i(E-\omega_\bfq)t} +\theta(-t)e^{i(E+\omega_\bfq)t}\Big) \nn
&\ =\lim_{a\rightarrow0}\int dt \, 
e^{-a^2 t^2/2}\,\Big(e^{i(E-\omega_\bfq)t} +e^{i(E+\omega_\bfq)t}\Big) 
-\int dt \, \Big(\theta(-t)e^{i(E-\omega_\bfq)t} +\theta(+t)e^{i(E+\omega_\bfq)t}\Big) \nn
&\ =\lim_{a\rightarrow0}\left( 2\pi\Delta_{a/\sqrt2}(E-\omega_\bfq) + 2\pi\Delta_{a/\sqrt2}(E+\omega_\bfq)\right)
- \left[ \frac{e^{i(E-\omega_\bfq)t}}{i(E-\omega_\bfq)} \right] ^0_{-\infty} 
- \left[ \frac{e^{i(E+\omega_\bfq)t}}{i(E+\omega_\bfq)} \right] _0^{+\infty} \nn
&\ = 2\pi\delta_{\rm c}(E-\omega_\bfq) + 2\pi\delta_{\rm c}(E+\omega_\bfq)
+ i\left(\frac1{E-\omega_\bfq} -  \frac1{E+\omega_\bfq}\right) \,.
\label{ImOmega>0}
\end{align}

Finally, we consider the `photon' which has a real energy $\omega_\bfq^a= \nu_\bfq$. 
In this case, we can follow the usual $-i\varepsilon$ trick to shift the mass square 
which results in the energy shift $\nu_\bfq\rightarrow\nu_\bfq-i\varepsilon$. Then, 
since $\Im(\nu_\bfq-i\varepsilon)<0$, this reduces to the above anti-ghost case and 
the result (\ref{ImOmega<0}) immediately gives 
\begin{align}
\int dt \, \Big(\theta(t)e^{i(E-\nu_\bfq+i\varepsilon)t} +\theta(-t)e^{i(E+\nu_\bfq-i\varepsilon)t}\Big) 
&= i\left( \frac1{E-\nu_\bfq+i\varepsilon} - \frac1{E+\nu_\bfq-i\varepsilon}\right) \,.
\label{ImOmega=0}
\end{align}
If we recall the well-known formula $1/(x \mp i\varepsilon) = \rmP(1/x) \pm i\pi\delta(x)$ 
(with $\rmP$ denoting principal value), 
this result can be rewritten into the form
\begin{equation}
\hbox{Eq.~(\ref{ImOmega=0})} = 
i\rmP\Big(\frac1{E-\nu_\bfq}\Big) - i\rmP\Big(\frac1{E+\nu_\bfq}\Big)
+\pi\delta(E-\nu_\bfq) +\pi\delta(E+\nu_\bfq) \,.
\end{equation}
It is interesting to note that this can further be rewritten into
\begin{equation}
= 
i\left( \frac1{E-\nu_\bfq-i\varepsilon} - \frac1{E+\nu_\bfq+i\varepsilon}\right) 
+2\pi\delta(E-\nu_\bfq) +2\pi\delta(E+\nu_\bfq) \,,
\label{ImOmega=01}
\end{equation}
which is the same result as the one which we would have obtained if 
we had used $+i\varepsilon$ trick to replace $\nu_\bfq \rightarrow\nu_\bfq+i\varepsilon$ and 
applied the formula (\ref{ImOmega>0}) for the ghost case. 
Since (\ref{ImOmega=0}) equals (\ref{ImOmega=01}), both tricks  replacing 
$\nu_\bfq$ by $\nu_\bfq \pm i\varepsilon$ lead to the same result.

%

Now, let us go back to the Eq.~(\ref{LHSof3.29}). 
We apply these formulas (\ref{ImOmega=0}), 
(\ref{ImOmega>0}) and (\ref{ImOmega<0}) to the $dt$ integration part
(\ref{Thetat-EXPizt}) %
for $\phi_a =$ `photon' $A$,\, ghost $\varphi$ and 
anti-ghost $\varphi^\dagger$, respectively, and  
divide out the total energy-momentum conservation factor
$i(2\pi)^4\delta^4(P'-P)$, setting $P=P'$ in the Eq.~(\ref{LHSof3.29}), 
to find the forward scattering transition amplitude
\begin{align}
&\bra{I}\hat{T}\ket{I} 
\big|_{\text{$s$-ch.}}\nn
&=
f^2\int d^3\bfq \,\delta^3(\bfq-\bfP) 
\biggl[
-\frac1{2\nu_\bfq}\Big(\frac1{P^0-\nu_\bfq+i\varepsilon} - \frac1{P^0+\nu_\bfq-i\varepsilon}\Big) 
\nn
&\hspace{1.5em}{} 
-\frac12\left( \frac1{-{P^0}^2+\bfq^2+M^2}  +i \frac{2\pi}{2\omega_\bfq} 
\left(\delta_{\rm c}(P^0-\omega_{\bfq}) 
+\delta_{\rm c}(P^0+\omega_{\bfq})\right)
+\frac1{-{P^0}^2+\bfq^2+{M^*}^2}\right)
\biggr] 
\nn
&=
f^2\biggl[
\frac1{P^2+\delta^2-i\varepsilon}
-\frac12\Bigl(
\frac1{P^2+M^2}+\frac1{P^2+{M^*}^2}\Bigr) 
-i \frac{\pi}{2\omega_\bfP}
\Bigl(\delta_{\rm c}(P^0-\omega_\bfP)+\delta_{\rm c}(P^0+\omega_\bfP)\Bigr)
\biggr] \,,\label{eq:3d-calculation}
\end{align}
where we have used $-{P_0}^2+{\omega^a_\bfq}^2=-P_0^2+\bfq^2+m_a^2=P^2+m_a^2$ 
with $m_a^2=( \delta^2, M^2, {M^*}^2)$ 
in the presence of $\delta^3(\bfP-\bfq)$.
We thus finally obtain the imaginary part of the $s$-channel forward 
transition amplitude,
\begin{equation}
2\Im\,\bra{I}\hat{T}\ket{I} 
\big|_{\text{$s$-ch.}}
= f^2 \bigg[
\frac{\pi}{\nu_\bfP}\,\delta(P^0-\nu_\bfP)
-\Re\Big[\frac{\pi}{\omega_\bfP}\,\delta_{\rm c}(P^0-\omega_\bfP)\Big] \bigg] 
\,,
\end{equation} 
where we have omitted the terms 
$ \delta(P^0+\nu_\bfP)$ and $\delta_{\rm c}(P^0+\omega_\bfP)$ since they vanish 
for $P^0>0$.\footnote{
This is because the complex delta function $\delta_{\rm c}(P^0+\omega_\bfP)$ has 
non-vanishing support only within 
$-\Re\omega_\bfP-\Im\omega_\bfP\leq P^0\leq-\Re\omega_\bfP+\Im\omega_\bfP$, but 
we have  $\Im \omega_{\bf0}< \Re\omega_{\bf0}$ for 
 $\Re M^2>0$ and $\Im M^2>0$ which 
leads to $\Im \omega_{\bfP}< \Re\omega_{\bfP}$,
implying $-\Re\omega_\bfP-\Im\omega_\bfP\leq P^0\leq-\Re\omega_\bfP+\Im\omega_\bfP
< 0$.
} 
This imaginary part exactly coincides with the RHS result (\ref{phiNorm}) 
divided by $(2\pi)^4 \delta^4(0)$,
that gives the 
$\phi$-production rate per unit space-time volume  directly calculated in the 
previous subsection \ref{sec:ghost-production}, 
thus confirming the optical theorem (\ref{OpticalTh}).

\subsubsection{Use of 4d-momentum form of the $\phi$ propagator}
\label{sec:4d-propagator}

Now we present the calculation of the $S$-matrix element 
(\ref{eq:S-1-0}) 
using the 4d-momentum form (\ref{4d-Prop}) of the propagator:
\begin{align}
&\bra{I'(\bfp'_1, \bfp'_2)}(S - 1)
\ket{I(\bfp_1, \bfp_2)}\big|_{\text{$s$-ch.}}\nn
&\quad {}=(if)^2 \int d^4x\,d^4y\,e^{-iP'x+iPy} \nn
&\qquad \times\int{d^3\bfq\over i(2\pi)^4} \left[ 
\int_R\frac{dq^0}{q^2+\delta^2}
-\frac12\bigg(
\int_C\frac{dq^0}{q^2+M^2} +\int_R\frac{dq^0}{q^2+{M^*}^2}
\bigg) \right] \,e^{iq(x-y)}, 
\label{eq:S-1-2}
\end{align}
where we recall that the $q^0$-integration contour for the ghost 
$\varphi$ is the much deformed $C$ drawn in Fig.~\ref{figC} (left), while those for 
`photon' $A$ and anti-ghost $\varphi^\dagger$ are along the real axis $R$. 
Then, performing the $d^4x$ and $d^4y$ integrations yields
\footnote{Note that the delta functions appearing here are generally {\it complex delta functions}, but the usual manipulation extracting the total 
energy-momentum conservation factor $(2\pi)^4\delta^4(P'-P)$ is possible as 
shown in the equation here and explained in the previous subsection.}
\begin{equation}
(2\pi)^4\delta_{\rm c}^4(P'-q)\cdot(2\pi)^4\delta_{\rm c}^4(q-P)
=(2\pi)^4\delta^4(P'-P)\cdot(2\pi)^4\delta_{\rm c}^4(q-P) \,,
\end{equation}
and after dividing out ($i$ times) the total momentum conservation factor, 
$i (2\pi)^4\delta^4(P'-P)$, and setting $P'=P$, 
we find the following expression for the $T$-matrix element for the 
forward scattering $s$-channel diagram:
\begin{eqnarray}
\bra{I}\hat{T}\ket{I} 
\big|_{\text{$s$-ch.}}
= f^2 
\left[ 
\int_R\frac{d^4q }{q^2+\delta^2-i\varepsilon}
-\frac12\bigg(
\int_C\frac{d^4q}{q^2+M^2} \!+\!\int_R\frac{d^4q}{q^2+{M^*}^2}
\bigg) \right] \delta^4_{\rm c}(q-P) \,.
\label{eq:3.44} 
\label{sch.Tamp}
\end{eqnarray}

Suppose that we perform the $d^4q$ integrations 
in this Eq.~(\ref{eq:3.44}) to  simply put $q^\mu$ equal to $P^\mu$ 
by naively using the complex delta function 
$\delta_{\rm c}(q-P)$, as if it were  just like the Dirac delta function. 
We would then immediately obtain
%
\footnote{The sum of the ghost and anti-ghost propagators 
in (\ref{WrongAns}) is
the real propagator for the fakeon \cite{Anselmi:2017ygm,Anselmi:2018kgz}.}
\begin{equation}
\bra{I}\hat{T}\ket{I} 
\big|_{\text{$s$-ch.}}
\ \overset{?}{=}\ f^2 
\left[ 
\frac1{P^2+\delta^2-i\varepsilon}
-\frac12\bigg(
\frac1{P^2+M^2} +\frac1{P^2+{M^*}^2}
\bigg) \right] \,,
\label{WrongAns}
\end{equation}
which is exactly the expression that one would write 
down if  {\it the usual Feynman rule in momentum space} is applied directly to 
the diagram Fig.~\ref{figN} (right), 
in which  the naive energy-momentum conservation at each vertex 
is tacitly assumed in advance.
This represents a very elementary, but very serious mistake 
which many have made, including Lee and Wick, and even many of those who were critical of their theory for other reasons. If Eq.~(\ref{WrongAns}) were correct, 
the ghost contributions, the second and third terms, add up to a real quantity 
and thus do not contribute to the imaginary part of the transition amplitude. 
Only the first `photon' part gives the following imaginary part (for the 
$s$-channel $P^0>0$):
\begin{eqnarray}
2\Im\,\bra{I}\hat{T}\ket{I} 
\big|_{\text{$s$-ch.}}
&= &f^2\frac{\pi}{\nu_\bfP}\delta(P^0-\nu_\bfP)\,,
\label{eq:photon}
\end{eqnarray}
where the term proportional to $\delta(P^0+\nu_\bfP)=0$ 
is suppressed. 
Essentially by this type of calculation and reasoning, Lee and Wick concluded 
that the 
complex ghosts are not produced from physical particle scattering.
%
%

The correct computation of the RHS of Eq.~(\ref{sch.Tamp}) should be performed as follows, 
noting that the naive use of the complex delta function 
$\delta_{\rm c}(q^0-P^0)$ is allowed when the integration variable $q^0$ 
remains real all the way along the contour since then $\delta_{\rm c}(q^0-P^0)$ 
is reduced to the usual Dirac delta function $\delta(q^0-P^0)$ with real 
argument $q^0-P^0$. 
Since this is already the case   
for the first term (`photon' part) and the third term (anti-ghost part), 
the problem lies in the second term (ghost part), for 
which the integration contour $C$ is much deformed one as depicted 
in Fig.~\ref{figC} (left)
on the complex $q^0$ plane. To avoid the moving imaginary part $\Im q^0$ on $C$, 
we can deform the contour $C$  according to 
\begin{equation}
\int_C dq^0 \  \rightarrow\ \bigg(\int_R + \int_{C(-\omega_\bfq)} - \int_{C(+\omega_\bfq)} \bigg) dq^0\,,
\label{deform2C2}
\end{equation}
as shown in Fig.~\ref{figC} (right), where $C(\omega)$ denotes an infinitesimal circle rotating anti-clockwise around 
the pole at $q^0=\omega$. Then, the first integral part $\int_R dq^0$ is again 
along the real axis, for which we can use the naive substitution rule 
$q^0 \rightarrow P^0$ also for the second term (ghost part).
This contribution 
together with those from the first and third terms in (\ref{eq:3.44}) reproduces 
Lee and Wick's amplitude (\ref{WrongAns}), which results from applying 
the usual Feynman rule with energy-momentum conservation assumed at 
each vertex. 

Here, however, we have two extra contributions from the two circles 
$C(\pm\omega_\bfq)$ which can be evaluated by the pole residues according 
to Cauchy's  theorem:\footnote{
It is important to note that the complex delta function $\delta_{\rm c}(q^0-P^0)$ 
is the $a\rightarrow0$ limit of the analytic (Gaussian) function $\Delta_a(q^0-P^0)$ which has 
no singularity in the relevant domain for the deformation performed in 
Eq.~(\ref{deform2C2}).}     
\begin{align}
- &\frac{f^2}2 \int d^3\bfq \, \delta^3(\bfq-\bfP) \bigg(\int_{C(-\omega_\bfq)} - \int_{C(+\omega_\bfq)} \bigg) dq^0 
\frac1{-{q^0}^2+\omega^2_\bfq}\delta_{\rm c}(q^0 - P^0) \nn
&= - \frac{f^2}2 \int d^3\bfq \, \delta^3(\bfq-\bfP)
\bigg(
\frac{2\pi i}{2\omega_\bfq}\Bigl( \delta_{\rm c}(\omega_\bfq-P^0) + \delta_{\rm c}(-\omega_\bfq-P^0) \Bigr)
\bigg)  \nn
&= - f^2 \frac{i\pi}{2\omega_\bfP}
\Bigl( \delta_{\rm c}(P^0-\omega_\bfP) + \delta_{\rm c}(P^0+\omega_\bfP) \Bigr)\,.
\label{ghost-cont}
\end{align}
Adding this to the Lee-Wick's amplitude (\ref{WrongAns}),  exactly 
reproduces the previous result (\ref{eq:3d-calculation}) 
obtained by using the 3d-momentum propagators.

At this stage we emphasize that in the 3d form of the propagators (\ref{varphiProp}),
the time propagation
of the positive and negative energy components are clearly distinguished 
thanks to $\theta(\pm t)$ function, 
where, in the case of the ghost, 
we mean  a positive or negative energy  with respect to its real part.
 As a result, there is no violation of causality even in the presence of the ghost.
 However,  the amplitude (\ref{WrongAns})
 will violate causality \cite{Coleman:1969xz,Donoghue:2019fcb,Donoghue:2021eto}
 (see also \cite{Lee:1969fy}), because
the integration  contour of the 4d form of the ghost and anti-ghost propagators 
(which inherits the causal information from the  3d form of the propagators) is completely   ignored.  
 
 \begin{figure}[ht]
\begin{center}
\hspace{0.7cm}
\includegraphics[width=7.2cm]{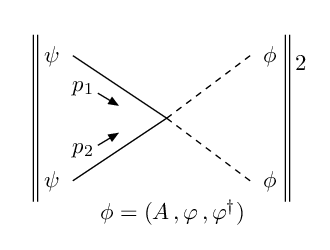}
\includegraphics[width=7.8cm]{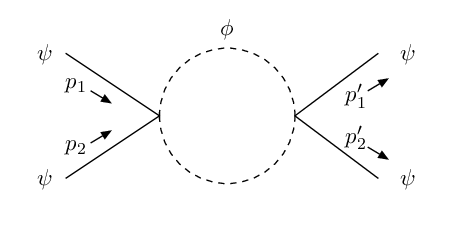}
\caption{Left: Graphical presentation of the norm (\ref{Norm-2phi}). Right: The one-loop amplitude for $ \psi+\psi\rightarrow\psi+\psi$}
\label{figX2}
\end{center}
\end{figure}

\section{Ghost pair production}\label{sec:pair-production}

The one-loop amplitude for $ \psi+\psi\rightarrow\psi+\psi$
shown in Fig.~\ref{figX2} (right)
has been representatively  considered to demonstrate 
the Lee-Wick prescription  \cite{Lee:1969fy,Lee:1969zze,Lee:1970iw} 
of how to integrate the loop momenta.
According to  their prescription, the imaginary part of 
the amplitude
should vanish for the parts where ghost and/or anti-ghost are 
propagating in the loop.
The optical theorem would then imply that 
the   production  of two ghost particles
by  a collision of two physical  particles, i.e.\
$ \psi+\psi\rightarrow B+B$\ ($B=(\varphi+\varphi^\dagger)/\sqrt2$)\  is forbidden. 
Here in this section we will examine their calculation in our field theory framework.
To this end, we consider the interaction Lagrangian
\begin{equation}
{\cal L}_{\rm int}(\psi,\phi)  =- \frac{f}2\, \psi^2\phi^2\,.\label{eq:Lint2}
\end{equation}

\subsection{Imaginary part calculation of the forward scattering amplitude}
\label{sec:forward scattering}

\subsubsection{Use of 3d-momentum form for the $\phi$ propagator}
\label{sec:3d-propagator2}

Since we are interested in ghost production, we focus 
only on the $s$-channel contribution to 
the one-loop amplitude for the scattering $\psi+\psi\rightarrow\psi+\psi$.  
With the interaction Lagrangian ${\cal L}_{\rm int}$ of (\ref{eq:Lint2}), 
the calculation proceeds quite in parallel to that in the previous section. 
As in Eq.~(\ref{eq:S-1-0}), we have at second order in the coupling $f$
\begin{align}
&\bra{I'(\bfp'_1, \bfp'_2)}(S - 1)
\ket{I(\bfp_1, \bfp_2)}
\big|_{\text{$s$-ch.}}
\Bigl( {}=i(2\pi)^4\delta^4(P'-P)\bra{I'(\bfp'_1, \bfp'_2)}\hat{T}\ket{I(\bfp_1, \bfp_2)}
\big|_{\text{$s$-ch.}} \Bigr)\nn
&\quad {}=\bra{I'(\bfp'_1, \bfp'_2)} \frac12 \Big(\frac{-if}2\Big)^2 
\rmT \int d^4x\,d^4y \,C_a(x^0)C_a(y^0)\,\psi^2(x)\phi^2(x)\,\psi^2(y)\phi^2(y)\ket{I(\bfp_1, \bfp_2)}
\big|_{\text{$s$-ch.}} \nn
&\quad {}=(if)^2 \int d^4x\,d^4y\,e^{-iP'x+iPy} \,e^{-a^2(x^0)^2-a^2(y^0)^2}
\,\Sigma(x-y) \,,\label{eq:S-1-3}
\end{align}  
with 
the initial $2\psi$ state 
$\ket{I(\bfp_1, \bfp_2)}$ defined in (\ref{Init2psi}) and 
$P=p_1+p_2\, ,P'=p'_1+p'_2$, where $C_a(x^0)$ is the adiabatic cutoff 
introduced in Eq.~(\ref{AdiabaticCutoff}) and $\Sigma(x-y)$ is given by
\begin{align}
\Sigma(x-y)&:= 2 \bra0 \rmT\,\phi(x)\phi(y)\ket0^2 
= 2 \sum_{a,b=1}^3 \eta_a\eta_b \Sigma_{ab}(x-y) \,, \nn 
\Sigma_{ab}(x-y)&= (-1)^{|a|+|b|} \bra0 \rmT \,\phi_a(x)\phi_a(y) \ket0
\bra0 \rmT \,\phi_b(x)\phi_b(y) \ket0\,.
\end{align}  
Here, $\phi_a = (A,\, \varphi,\, \varphi^\dagger)$ are the three component fields 
in $\phi= A + (\varphi+\varphi^\dagger)/\sqrt{2}$ with the weight factor 
$\eta_a=( 1,\, -1/2,\, -1/2)$ and norm sign $(-1)^{|a|}=( +1,\, -1,\,-1)$. 
Inserting the 3d momentum expression in 
Eq.~(\ref{3d-Prop}) for the propagator 
$\bra0 \rmT \,\phi_a(x)\phi_a(y) \ket0$, we find    
a compact form for $\Sigma_{ab}$:
\begin{align}
\Sigma_{ab}(x-y)
&=
\int\frac{d^3\bfq \,d^3 \bfq'}
  {(2\pi)^6 2\calE^a_\bfq 2\calE^b_{\bfq'}} \Big(
  \theta(x^0-y^0) e^{i (\what q_a+\what q'_b) (x-y)}+ 
   \theta(y^0-x^0) e^{-i(\what q_a+\what q'_b)(x-y)}\Big)\,\label{Sigma}\,,
\end{align}
where $\calE^a_\bfq$ and $\what q_a^\mu$ are the energy and the on-shell 
4-momentum of $\phi_a$ defined before in Eq.~(\ref{EnergyOnShellMomentum}).
%

Noting the similarity between Eqs.~(\ref{eq:S-1-0}) and (\ref{eq:S-1-3}), 
and recalling the procedure to obtain Eq.~(\ref{LHSof3.29}), 
we perform the $d^3\bfx d^3\bfy$ integrations in (\ref{eq:S-1-3}),
yielding three dimensional delta functions $(2\pi)^3\delta^3(\bfq+\bfq'-\bfP)$ and 
$(2\pi)^3\delta^3(\bfq+\bfq'-\bfP')$. This then allows us to carry out
the integration over $\bfq'$ to write $\bfq'=-\bfq+\bfP
=-\bfq+\bfP'$.
\footnote{The second term in (\ref{Sigma}), i.e., \ the one proportional to
$\theta(-t)$, actually gives  $\bfq'=-\bfq-\bfP
=-\bfq-\bfP'$, but we make the change $\bfq\to -\bfq$ which
implies $-\bfq-\bfP\to \bfq-\bfP$. We then use the fact that $\nu_{\bfq}=
\nu_{-\bfq}$ and $\omega_{\bfq}=\omega_{-\bfq}$.
}
Then, as we have done in (\ref{CM-time}),  we use the CM and relative time
$(T=(x^0+y^0)/2,\ t=x^0-y^0)$
to perform the time integration and obtain
\begin{align}
&\bra{I'(\bfp'_1, \bfp'_2)}(S - 1)
\ket{I(\bfp_1, \bfp_2)}
\big|_{\text{$s$-ch.}}\nn
&=-f^2(2\pi)^3\delta^3(\bfP-\bfP')
 \Big(\int dT\,
e^{i(P^0-P'^0)T}e^{-2 a^2 T^2}\Big)\,\nn
&\times\int\frac{d^3\bfq}{(2\pi)^3}\,\sum_{a,b}
\frac{2\eta_a\eta_b}{2\calE^a_\bfq\, 2\calE^b_{\bfP-\bfq}}
\int dt\,e^{-a^2 t^2/2}\left\{
\theta(t)\,e^{i(\bar{P^0}-\bar q^0_{ab})t}
+\theta(-t)\,e^{i(\bar{P^0}+\bar q^0_{ab})t}
\right\}\,,\label{eq:S-1-4}
\end{align}
where 
$\bar{P^0}=(P^0+P'^{0})/2=(p^0_1+p^0_2+p'^{0}_{1}+
p'^{0}_{2})/2$ and $\bar q^0_{ab}=\omega^a_\bfq+\omega^b_{\bfP-\bfq}$, and
the $T$ integration gives $(2\pi)\delta(P^0-P'^0)$ in the $a\to 0$ limit. 
When applying the formulas to carry out the $t$ integration with either (\ref{ImOmega>0}),
(\ref{ImOmega<0}), or (\ref{ImOmega=0})
and $\theta(\pm t)$, we must classify the cases when the imaginary part of 
$\bar q^0_{ab}=\omega^a_\bfq+\omega^b_{\bfP-\bfq}$ becomes positive or negative, 
which is a bit complicated  task. In order to avoid such an inessential 
complication for the present issue of ghost production, we discuss 
the problem in the CM frame and set $\bfP={\bf 0}$ henceforth. 
Then, the equality $\omega^a_{\bfP-\bfq}=\omega^a_\bfq$ holds so that we have 
\begin{equation}
\begin{array}{cccl}
\Im \bar q^0_{ab}=\Im(\omega_\bfq^a+\omega_\bfq^b) && (\phi_a,\,\phi_b)\hbox{\ or\ } 
(\phi_b,\,\phi_a) &\text{Applied formula}\\ \hline
\Im \bar q^0_{ab}=0  &:&(A,\,A), \ (\varphi,\,\varphi^\dagger) & ~~~\text{Eq.}(\ref{ImOmega=0})   \\ 
\Im \bar q^0_{ab}>0  &:&(A,\,\varphi),\ (\varphi,\,\varphi) & ~~~\text{Eq.}(\ref{ImOmega>0})           \\
\Im \bar q^0_{ab}<0  &:&(A,\,\varphi^\dagger),\ (\varphi^\dagger,\,\varphi^\dagger) & ~~~\text{Eq.}(\ref{ImOmega<0}) \,.
\end{array}\label{CasesTable}
\end{equation}
Applying the indicated formula for the $dt$ integrations in (\ref{eq:S-1-3}), 
dividing out the factor $i (2\pi)^4 \delta^4(P' - P)$ and setting $P=P'$, 
we obtain the forward scattering $\hat T$ amplitude 
\begin{align}
&\bra{I(\bfp_1, \bfp_2)}\hat T\ket{I(\bfp_1, \bfp_2)}\big|_{\text{$s$-ch.}} \nn
&=\frac{f^2}{2}
\int\! \frac{d^3 \bfq}{(2\pi)^3}\ 
\Biggl\{
\frac1{\nu_\bfq^2}\,\frac{ 4\nu_\bfq}{(2\nu_\bfq)^2 -{P^0}^2-i\varepsilon} \nn
& 
\qquad {}-\frac1{\nu_\bfq \omega_\bfq}\,
\frac{ 2(\nu_\bfq+\omega_\bfq)}{(\nu_\bfq + \omega_\bfq) ^2 -{P^0}^2}
-i\frac{2\pi}{\nu_\bfq\omega_\bfq}\,
\delta_\rmc(P^0-\nu_\bfq-\omega_\bfq) 
-\frac1{\nu_\bfq\omega^*_\bfq}\,
\frac{ 2(\nu_\bfq+\omega^*_\bfq)}{(\nu_\bfq + \omega^*_\bfq) ^2 -{P^0}^2}\nn
&\qquad {}+\frac1{2\omega_\bfq \omega^*_\bfq}\,
\frac{ 2(\omega_\bfq+\omega^*_\bfq)}{(\omega_\bfq + \omega^*_\bfq) ^2 -{P^0}^2-i\varepsilon} \nn
&\qquad  {}+\frac1{4\omega_\bfq^2}\,
\frac{ 4\omega_\bfq}{ (2\omega_\bfq)^2 -{P^0}^2} 
+i\frac{2\pi}{4\omega_\bfq^2}\, 
\delta_\rmc(P^0-2\omega_\bfq) 
+\frac1{4{\omega^*_\bfq}^2}\,
\frac{4\omega^*_\bfq}{ (2{\omega^*_\bfq})^2 -{P^0}^2}
\Biggr\}\,,\label{eq:S-1-5}
\end{align}
where we have omitted the terms $\delta_\rmc(P^0+2\omega_\bfq)$ and 
$\delta_\rmc(P^0+\nu_\bfq+\omega_\bfq)$ which vanish for the $s$-channel $P^0>0$ 
because of (\ref{LowerUpperThreshold}), 
and used $\bar{P}^0=(P^0+{P'}^0)/2=P^0$.
Note that we applied Eq. (\ref{ImOmega=0}) for the $(\varphi,\varphi^\dag)$ loop
because ${\rm Im} \,(\omega_\bfq+\omega^*_\bfq)=0$.
This result (\ref{eq:S-1-5}) for the one-loop $\hat T$ amplitude consists of fraction 
and complex delta function pieces. All the fraction parts other than that from the
$(\varphi, \varphi^\dagger)$-loop (containing $(\omega_\bfq + \omega^*_\bfq) ^2$) are    
identical with the result of Lee \cite{Lee:1969zze}. However, the complex delta functions 
 and the fraction  part from the $(\varphi, \varphi^\dagger)$-loop with $-i\varepsilon$,
which contribute to the imaginary part corresponding to the ghost 
production rate, are all {\it absent} in Lee's computation \cite{Lee:1969zze}.

From our result (\ref{eq:S-1-5}), the correct imaginary part of 
the forward scattering $\hat T$ amplitude is found to be 
\begin{align}
&2\Im \bra{I(\bfp_1, \bfp_2)}\hat T\ket{I(\bfp_1, \bfp_2)}\big|_{\text{$s$-ch.}} \nn
&=2 f^2
\int\! \frac{d^3 \bfq}{(2\pi)^3}\ 
\biggl\{
\frac{\pi}{2\nu_\bfq^2}\,
 \delta(P^0-2\nu_\bfq)  
+\frac{\pi}{4\omega_\bfq \omega^*_\bfq}\, 
\delta(P^0-\omega_\bfq - \omega^*_\bfq) \nn
& \hspace{7em}{}
- \Bigl(
\frac{\pi}{2\nu_\bfq\omega_\bfq}\delta_\rmc(P^0-\nu_\bfq-\omega_\bfq) 
+ \frac{\pi}{2\nu_\bfq\omega^*_\bfq}\delta_\rmc(P^0-\nu_\bfq-\omega^*_\bfq)
\Bigr) \nn
&\hspace{7em}{}+\Bigl(
\frac{\pi}{8\omega_\bfq^2}\,\delta_\rmc(P^0-2\omega_\bfq) 
+\frac{\pi}{8{\omega^*_\bfq}^2}\,\delta_\rmc(P^0-2\omega^*_\bfq) \Bigr) 
\bigg\}\ .
\label{eq:Im<T>}
\end{align}
Again, we have omitted all the terms of the form 
$\delta_\rmc(P^0+ \omega^a_\bfq+\omega^b_\bfq)$ which vanish when $P^0>0$.
Needless to say that only the first term (`photon-photon' term) is present  in Lee's computation. 

\subsubsection{Use of 4d-momentum expression of the $\phi$ propagator}
\label{sec:4d-propagator2}

Here we present the calculation of the matrix element 
(\ref{eq:S-1-3}) by using the 4d-momentum form of the $\phi$ propagator
(\ref{4d-Prop}). We use $\Sigma_{ab}(x-y)$ in Eq. (\ref{Sigma})
 in the 4d form
\begin{align}
\Sigma_{ab}(x-y) &= \int_{C_a}\frac{d^4 q}{i(2\pi)^4}\int_{C_b}
\frac{d^4 q'}{i(2\pi)^4}\,e^{i(x-y)(q+q')}\,
\frac{1}{q^2+m_a^2}\,\frac{1}{q'^2+m_b^2}\,,
\end{align}
where $C_a= (R,\, C,\, R)$ for $\phi_a=(A,\,\varphi,\,\varphi^\dagger)$,
respectively,
denotes the $q^0$-integration contour and $C_b$ denotes the $q'^0$-integration 
contour. 
Note also that we should put $-i\varepsilon$ for the `photon' (or real mass particle) case
so that the mass squared $m^2_a$ here is understood as 
$m^2_a= (\delta^2-i\varepsilon,\, M^2,\,{M^*}^2)$. 

%
%

The $d^4x$ and $d^4y$ integrations in Eq.~(\ref{eq:S-1-3}) give in the $a\rightarrow0$ 
limit
\begin{eqnarray}
& &(2\pi)^8\,\delta^3(\bfq+\bfq'-\bfP')\,\delta^3(\bfP-\bfq-\bfq')
\delta_\rmc(q^0+{q'}^0-{P'}^0)\,\delta_\rmc(P^0-q^0-{q'}^0)\nn
& &=(2\pi)^8\,\delta^4(P-P')
\delta^3(\bfP-\bfq-\bfq')
\delta_\rmc(P^0-q^0-{q'}^0)\,.
\end{eqnarray}
These delta functions, other than the last one, are all the usual Dirac 
functions. Using these usual delta functions we can carry out the 
$\bfq'$ integration,
divide out the factor $i (2\pi)^4 \delta^4(P' - P)$ and set $P=P'$ 
to obtain the forward scattering $\hat T$ amplitude 
\begin{align}
&\bra{I(\bfp_1, \bfp_2)}\hat T\ket{I(\bfp_1, \bfp_2)}\big|_{\text{$s$-ch.}} \nn
&=f^2 \sum_{a,b}\eta_a\eta_b \int\frac{d^3\bfq}{(2\pi)^3} 
\int_{C_a}\frac{dq^0}{2\pi i} 
\times\int_{C_b}dq'^0\,
\frac{1}{{\omega^a_\bfq}^2-{q^0}^2}\,
\frac{1}{{\omega^b_{\bfq'}}^2-{q'^0}^2}\,\delta_\rmc(P^0-q^0-q'^0)
,\label{eq:S-1-6}
\end{align}
where $\bfq'=\bfP-\bfq$ is understood, and 
$\omega_\bfq^a=(\nu_\bfq,\ \omega_\bfq,\ \omega^*_\bfq)$. 
We should however emphasize that {\it the $q^0$- or 
$q'^0$-integration 
using the complex delta function $\delta_\rmc(P^0-q^0-q'^0)$ is 
very non-trivial} because the argument $P^0-q^0-q'^0$ is complex when 
$\phi_a$ or $\phi_b$ is the ghost $\varphi$. In such cases, we have to make 
a shift of the integration contour to make $q^0+q'^0$ real so as to reduce 
the complex delta function $\delta_\rmc$ to the usual 
Dirac delta function $\delta$, thus allowing it to pick up the pole singularities of 
the propagators. 

We may then perform this task for all the $3\times3=9$ cases of combination 
of two field propagators in Eq.~(\ref{eq:S-1-6}) systematically. This task is 
straightforward but becomes slightly tedious and a bit lengthy, so we 
move the calculation to the Appendix. There we actually obtain exactly 
the same 
result as the above Eq.~(\ref{eq:S-1-5}) obtained by using the 3d form of 
propagators. 

Here, however, we should concisely explain the reason 
why the usual Feynman 
rule expression (\ref{eq:wrong}) with energy-momentum conservation used 
in advance is {\it wrong} and does not follow from the correct 
expression (\ref{eq:correct}), as announced in the Introduction. 
The proofs for this are given for many concrete cases 
in the Appendix, however, since the essential point 
 is buried in the lengthy calculation, it would be better to recapitulate it here. 

We start with the correct expression (\ref{eq:correct}) in the Introduction, 
which 
reads, after performing the 3d $\bfk$ integration by using the usual 
Dirac delta function $\delta^3(\bfk+\bfq-\bfp)$,
\begin{align}
\int d^3\bfq\int_Cdq^0\int_Rdk^0 \, \frac1{\omega_\bfq^2-{q^0}^2}\,
\frac1{E_{\bfp-\bfq}^2 -{k^0}^2-i\varepsilon}\delta_\rmc(k^0+q^0-p^0)\,,
\label{eq:start}
\end{align}
where $\omega_\bfq=\sqrt{\bfq^2+M^2}$ and $E_\bfk=\sqrt{\bfk^2+\mu^2}$ are the 
energies of the complex ghost $\varphi$ and physical particle $\psi$, 
respectively. We have written the $k^0$ integration contour as $R$ (real axis)
explicitly while putting the usual $-i\varepsilon$ to indicate how to avoid the 
pole on the real axis.
This expression corresponds to the case for $\phi_a=\varphi$ (ghost) 
and $\phi_b = \psi$ (physical particle in place of the photon $A$) in 
Eq.~(\ref{eq:S-1-6}). 

Now the problem is the fact that $q^0$ is complex since 
the $q^0$-integration contour $C$ is the 
much deformed contour in the complex $q^0$ plane as shown in Fig.~1. 
So, when using the complex delta function $\delta_\rmc(k^0+q^0-p^0)$ for 
$k^0$ integration, we need to make the argument $k^0+q^0$ real by 
shifting the integration variable $k^0$ in order to reduce the 
$\delta_\rmc(k^0+q^0-p^0)$ to the usual Dirac delta function.  
To do so, it is better to make the imaginary part $\Im q^0$ not run.
So we change the $q^0$-integration contour $C$ in Fig.~1 (left) to the 
contour Fig.~1 (right) as we did in Eq.~(\ref{deform2C2}) for the single 
ghost production case. 
Then, the $q^0$-integration part in Eq.~(\ref{eq:start}) can be rewritten 
into
\begin{equation}
\int_C dq^0 \frac1{\omega_\bfq^2-{q^0}^2}\delta_\rmc(k^0+q^0-p^0)
=\int_R dq^0 \frac1{\omega_\bfq^2-{q^0}^2}\delta(k^0+q^0-p^0)
+\frac{2\pi i}{2\omega_\bfq}\sum_{\pm}\delta_\rmc(k^0\pm\omega_\bfq-p^0)\,.
\label{eq:A1}
\end{equation}
The last $\sum_{\pm}$ terms are the contributions from $\int_{C(\pm\omega_\bfq)}dq^0$ 
around the poles at $q^0=\pm\omega_\bfq$. Note that the complex delta function 
$\delta_\rmc(k^0+q^0-p^0)$ has become the usual Dirac delta function 
$\delta(k^0+q^0-p^0)$ in the first term on the RHS
since $q^0$ remains real on the contour $R$.
Inserting this and performing the trivial $k^0$ integration using this Dirac 
delta function in the first term, the Eq.~(\ref{eq:start}) becomes
\begin{align}
\hbox{Eq.~(\ref{eq:start})}
&=\int d^3\bfq\bigg\{
\int_Rdq^0 \frac1{\omega_\bfq^2-{q^0}^2}\,
\frac1{E_{\bfp-\bfq}^2 -(p^0-q^0)^2-i\varepsilon} \nn
&\hspace{5em}+ 
\frac{2\pi i}{2\omega_\bfq}\sum_{\pm}\int_R dk^0 \,\delta_\rmc(k^0\pm\omega_\bfq-p^0)
\frac1{E_{\bfp-\bfq}^2 -{k^0}^2-i\varepsilon}\bigg\}.
\label{eq:start2}
\end{align}
The $k^0$-integration for the second two ($\pm$) terms is non-trivial. 
To make the 
arguments $k^0\pm\omega_\bfq-p^0$ of $\delta_\rmc$ real there, 
we shift the $k^0$-integration contour 
from the real axis $R$ to the contour $R(\mp \omega_\bfq)$, respectively. 
Here $R(z_0)$ with a complex $z_0$ 
generally indicates the horizontal contour parallel to the real axis and 
passing the point $z_0$.  
However, the point is that this shift of the $k^0$-integration contour 
from $R$ to $R(\mp\omega_\bfq)$ does not change the integral,
 only if there are no singularities 
in the $k^0$ integrand in the $k^0$-domain between the 
two contours $R$ and $R(\mp \omega_\bfq)$ 
(provided that the contributions from the vertical passes at $\pm\infty$ between 
them vanish as is the case here).
In this rectangular domain, however, the present integrand 
$1/ (E^2_{\bfp-\bfq}-{k^0}^2-i\varepsilon)$ actually has  a pole 
at $k^0 = \pm(E_{\bfp-\bfq}-i\varepsilon)$, respectively, so that  
this shift also picks up the contribution from the 
poles and we have
\begin{equation}
\int_R dk^0\ \frac{\delta_\rmc(k^0\pm\omega_\bfq-p^0)}{E_{\bfp-\bfq}^2 -{k^0}^2-i\varepsilon}
= \left(\int_{R(\mp\omega_\bfq)}  \mp
\int_{C\bigl(\pm(E_{\bfp-\bfq}-i\varepsilon)\bigr)}\right)dk^0\ 
\frac{\delta_\rmc(k^0\pm\omega_\bfq-p^0)}{E_{\bfp-\bfq}^2 -{k^0}^2-i\varepsilon} \,.
\end{equation}
with $C\bigl(\pm(E_{\bfp-\bfq}-i\varepsilon)\bigr)$ denoting the infinitesimal circle 
surrounding the pole $k^0 = \pm(E_{\bfp-\bfq}-i\varepsilon)$ anti-clockwise. 
Note that $k^0\pm\omega_\bfq$ is real on the contour $R(\mp\omega_\bfq)$ 
for the first term so that $\delta_\rmc(k^0\pm\omega_\bfq-p^0)$ reduces to the 
Dirac delta function and hence the $k^0$ integration becomes trivial. 
Noting also that the integral for the second term is 
evaluated by the Cauchy theorem, we find 
\begin{align}
&\int_R dk^0 \,\delta_\rmc(k^0\pm\omega_\bfq-p^0)
\frac1{E_{\bfp-\bfq}^2 -{k^0}^2-i\varepsilon} \nn
& \quad =
\frac1{E_{\bfp-\bfq}^2 -(p^0 \mp \omega_\bfq)^2}
+\frac{2\pi i}{2E_{\bfp-\bfq}}\delta_\rmc\bigl(\pm(E_{\bfp-\bfq}+\omega_\bfq)-p^0\bigr)\,.
\end{align}
Substitution of this into the second term in Eq.~(\ref{eq:start2}) gives 
\begin{align}
\hbox{Eq.~(\ref{eq:start})}
&=\int d^3\bfq\bigg\{
\int_Rdq^0 \frac1{\omega_\bfq^2-{q^0}^2}\,
\frac1{E_{\bfp-\bfq}^2 -(p^0-q^0)^2-i\varepsilon} \nn
&\hspace{5em}+ 
\sum_{\pm}\frac{2\pi i}{2\omega_\bfq}
\frac1{E_{\bfp-\bfq}^2 -(p^0\mp\omega_\bfq)^2} \nn
&\hspace{5em}+\sum_{\pm}
\frac{2\pi i}{2\omega_\bfq}
\frac{2\pi i}{2E_{\bfp-\bfq}}\delta_\rmc\bigl(\pm(E_{\bfp-\bfq}+\omega_\bfq)-p^0\bigr)
\bigg\}.
\label{eq:start3}
\end{align}
We immediately notice that the first plus second terms in fact 
give the original usual Feynman rule expression (\ref{eq:wrong}) 
with normal energy-momentum conservation used:
\begin{equation}
\int_C d^4q \, \frac1{q^2+M^2}\,\frac1{(p-q)^2+\mu^2}\ . 
\end{equation}
Note that the integration contour is $C$. 
So we find that this naive Feynman rule misses the extra contributions 
expressed by the third term in Eq.~(\ref{eq:start3}).

If combined with the contribution from the $\varphi^\dagger$- 
$\psi$ loop, the naive Feynman rule gives only a 
real amplitude as Lee \cite{Lee:1969zze} showed even if the deformed contour $C$ is 
used correctly. The imaginary parts solely come from the 
extra third term in Eq.~(\ref{eq:start3}), which originates from the complex delta
function.

\subsection{Direct calculation of two $\phi$ production}
\label{sec:direct-production}
Because of the optical theorem  (\ref{OpticalTh}), the  imaginary  part of (\ref{eq:S-1-3})
should be equal to  the norm of the intermediate state that also enters 
 the unitarity sum:
 \begin{equation}
\Big|\!\Big| (S-1)\, \ket{I(\bfp_1, \bfp_2)}\Big|\!\Big|^2=
\bra{I(\bfp_1, \bfp_2)}\, (S-1)^\dagger(S-1)\, \ket{I(\bfp_1, \bfp_2)}\,,
\end{equation}
where, to the first order in the coupling $f$ of ${\cal L}_{\rm int}$ in 
Eq.~(\ref{eq:Lint2}), 
\begin{align}
& (S-1)\, \ket{I(\bfp_1, \bfp_2)} \nn
&= \frac{-if}{2} \int d x^0 C_a(x^0)
\int d^3 \bfx\,\psi(x)^2 \phi(x)^2\, \ket{I(\bfp_1, \bfp_2)}\nn
&=-if\int dx^0 \int d^3 \bfx  \,C_a(x^0)\,
e^{i(p_1^0+p_2^0-\omega^a_{\bfq}-\omega^b_{\bfq'})x^0} \nn
&\times\sum_{a,b}\sqrt{|\eta_a\eta_b|}\int\frac{d^3 \bfq}{\sqrt{(2\pi)^3}}\int\frac{d^3 \bfq'}{\sqrt{(2\pi)^3}}
\,e^{-i(\bfp_1+\bfp_2-\bfq-\bfq')\bfx}\,
\,\frac{\phi_a^\dag(\bfq) }{\sqrt{2 \omega^a_\bfq}}
\,\frac{\phi_b^\dag(\bfq') }{\sqrt{2 \omega^b_{\bfq'}}}\,\ket0\,,
\label{eq:M4-1}
\end{align}
and $C_a(x^0)$ is the adiabatic cutoff given in (\ref{AdiabaticCutoff}).
In this expression (\ref{eq:M4-1})
we have omitted the other states corresponding to the disconnected diagrams which 
are irrelevant to the present discussion.  
Further,  $\omega^a_{\bfq}$, $|\eta_a|$ and $\phi_a^\dag(\bfq)$ with index $a$ running 
over three component fields $\phi_a=( A,\, \varphi, \, \varphi^\dagger)$ are given by
\begin{equation}
\omega^a_\bfq= (\nu_\bfq,\,\omega_\bfq,\,\omega^*_\bfq), \ \ 
|\eta_a| = (1,\, 1/2,\,1/2) \ \ \hbox{and} \ \ 
\phi^\dagger_a(\bfq)=( a^\dagger(\bfq),\, \beta^\dagger(\bfq),\, \alpha^\dagger(\bfq)\,)\,,
\end{equation}
respectively. 
The integration over $\bfx$ gives a three-dimensional delta function
$(2\pi)^3\,\delta^3(\bfP-\bfq-\bfq')$ with $\bfP=\bfp_1+\bfp_2$, which can be 
used to perform the integration over $\bfq'$. 
Then, performing also the $x^0$ integration by using the definition of the 
complex delta function (\ref{Delta-a}), we obtain in the $a\rightarrow0$ limit 
\begin{align}
& (S-1)\, \ket{I(\bfp_1, \bfp_2)} \nn
&=-if\int d^3\bfq \sum_{a,b}\sqrt{|\eta_a\eta_b|}\,
\frac{\phi_a^\dagger(\bfq)}{\sqrt{2\omega^a_\bfq}}\,
\frac{\phi_b^\dagger(\bfq')}{\sqrt{2\omega^b_{\bfq'}}}\,\ket0 \,
2\pi\delta_\rmc(\omega^a_\bfq+\omega^b_{\bfq'}-P^0) ~~~\mbox{with}~~\bfq'=\bfP-\bfq\,.
\label{eq:M4-2}
\end{align}
We note that 
\begin{equation}
\sum_a \sqrt{|\eta_a|}\,
\frac{\phi_a^\dagger(\bfq)}{\sqrt{2\omega^a_\bfq}}\,\ket0  = 
\frac1{\sqrt{2\what{q}^0}}\,\phi^\dagger(\bfq)\,\ket0 
\label{phi-state3}
\end{equation}
is just the superposition of one-particle states created by the field 
$\phi(x) = A+ (\varphi +\varphi^\dagger)/\sqrt2$ at time $x^0=0$ as 
introduced in Eq.~(\ref{phi-state}),  whose norm 
has been already calculated when obtaining
Eq.~(\ref{phiNorm}). In place of the norm, the inner-product is also 
computed in the same way as  
\begin{align}
\bra0 \phi(\bfk)\frac1{\sqrt{2\what{k}^0}}\,
\frac1{\sqrt{2\what{q}^0}}\,\phi^\dagger(\bfq)\,\ket0  
&= \delta^3(\bfk-\bfq)\left\{
\frac1{2\nu_\bfq} - \half \Big( \frac1{2\omega_\bfq} + \frac1{2\omega^*_\bfq} \Big)
\right\} 
=
\delta^3(\bfk -\bfq) 
\sum_a \eta_a \frac 1{2\omega^a_\bfq}\,.
\label{phiNorm3}
\end{align}
Since the state appearing here 
(\ref{eq:M4-2}) is just a two-particle (tensor product) state of 
(\ref{phi-state3}), 
the norm of (\ref{eq:M4-2}) can immediately be found from
Eq.~(\ref{phiNorm3}) to read   
\begin{equation}
\Big|\!\Big| (S-1)\, \ket{I(\bfp_1, \bfp_2)}\Big|\!\Big|^2
= (2\pi)^4 \delta^4(0)\nn
 \ 2 f^2 \,
\sum_{a,b} \eta_a\eta_b\,
\int\frac{d^3 \bfq}{(2\pi)^3(4\omega^a_\bfq\,\omega^b_{\bfP-\bfq})}\,
2\pi\delta_\rmc(\omega^a_\bfq+\omega^b_{\bfP-\bfq}-P^0)\,.
\label{eq:Norm4}
\end{equation}
Dividing out the factor $(2\pi)^4\delta^4(0)$ and writing the sum over $a, b$ 
explicitly, we find the following expression for the norm of the 
produced state ${\hat T} \ket{I}$ in the CM frame ($\bfP={\bf0}$):
\begin{align}
&\Big|\!\Big| \hat T\,\ket{I(\bfp_1, \bfp_2)}\Big|\!\Big|^2 \nn
&= 2f^2
\int\frac{d^3 \bfq}{(2\pi)^3}
\biggl\{
\frac{\pi}{2\nu_\bfq^2}\,
\delta(2\nu_\bfq-P^0)  
+\frac{\pi}{4\omega_\bfq\,\omega^*_\bfq}\,
\delta(\omega_\bfq+\omega^*_\bfq-P^0) 
\nn
&\hspace{7em}{}
-\Bigl(\frac{\pi}{2\nu_\bfq\,\omega_\bfq}\,
\delta_\rmc(\nu_\bfq+\omega_\bfq-P^0) 
+\frac{\pi}{2\nu_\bfq\,\omega^*_\bfq}\,
\delta_\rmc(\nu_\bfq+\omega^*_\bfq-P^0) 
\Bigr)\nn
&\hspace{7em}{}
+\Bigl(\frac{\pi}{8\omega_\bfq^2}\,
\delta_\rmc(2\omega_\bfq-P^0) 
+\frac{\pi}{8{\omega^*_\bfq}^2}\,
\delta_\rmc(2\omega^*_\bfq-P^0)
\Bigr) 
\bigg\}\,.
\label{Norm-2phi}
\end{align} 
We see that this result giving the RHS of the optical theorem (\ref{OpticalTh}),
exactly coincides with
the imaginary part of the forward scattering amplitude given in 
Eq.~(\ref{eq:Im<T>}) which we 
calculated in the previous subsection.   
The first term in the first line gives 
the production probability of two `photon' $A$-$A$ which is of course 
positive as usual. 
The second term in the first line and the two terms in the third line 
give the production probability of two ghosts via 
$\varphi$-$\varphi^\dagger$, $\varphi$-$\varphi$ and $\varphi^\dagger$-$\varphi^\dagger$, 
respectively, and are positive due to the two negative norm particles. 
The two terms in the second line give a negative probability for the 
production of a `photon' and a ghost, $A$-$\varphi$ and $A$-$\varphi^\dagger$.
These probabilities are all, except for the normal two 'photon' case, 
proportional to the complex delta function. We have already shown in Section 
\ref{sec:single-ghost} 
that the complex delta function is well-defined and non-vanishing, so these probabilities for two ghost particles are clearly non-vanishing 
and violate the physical particles' unitarity. 
Since these probabilities contain the 3d momentum integration 
$d^3\bfq$ of the complex delta function, this shows a new interesting aspect 
of this complex delta function, so that we may now discuss more explicitly 
the production probability of two ghosts, 
$\varphi$-$\varphi^\dagger$, $\varphi$-$\varphi$ and $\varphi^\dagger$-$\varphi^\dagger$.

We thus have explicitly confirmed that 
the optical theorem  (\ref{OpticalTh}) is satisfied 
for our scattering process in the lowest non-trivial order. 
This theorem is a trivial identity which directly follows 
from the unitarity of the 
Dyson's $S$-matrix as a result of the hermiticity of our interaction 
Lagrangian. So although the confirmation itself of the theorem 
does not have so important  meaning, it gives a useful consistency 
check for the validity of our computations. 
 

\subsection{Explicit evaluation of the  ghost production probability} 
\label{sec:explicit-evaluation}

\subsubsection{Two ghosts production}
\label{sec:two-ghosts} 

Let us now investigate more explicitly 
the production probability for two ghosts,
$\varphi$-$\varphi^\dagger$, $\varphi$-$\varphi$ and $\varphi^\dagger$-$\varphi^\dagger$, 
which were given  in 
the second term of the first line and the two terms in the third line
in Eq.~(\ref{Norm-2phi}), 
respectively. 

First consider the ghost--anti-ghost $\varphi$-$\varphi^\dagger$ pair 
production, whose probability is given by 
\begin{equation}
f^2
\int\frac{d^3 \bfq}{(2\pi)^3}
\frac{\pi}{2\omega_\bfq\,\omega^*_\bfq}\,
\delta(\omega_\bfq+\omega^*_\bfq-P^0)\,. 
\label{varphi-Conjvarphi}
\end{equation}
This pair production is very special  since the total 
energy $\omega_\bfq+\omega^*_\bfq$ of the complex ghost pair can be exceptionally
{\it real} in the CM frame with $\bfP=0$.
The delta function $\delta(\omega_\bfq+\omega^*_\bfq-P^0)$ here is the 
Dirac delta function implying the usual energy conservation which 
determines the magnitude of momentum $|\bfq|\equiv q$ of produced ghost as
\begin{equation}
q^2 = E^2 - \Bigl(m^2 + \frac{\gamma^4}{4E^2}\Bigr) =: q^2_{\rm CM}(E)\,,
\end{equation}
where the total energy is $P^0=2E$. It is interesting to note that the 
imaginary part $\gamma^2$ of the complex ghost mass squared
 $M^2 = m^2+i\gamma^2$ 
effectively contributes $\gamma^4/4E^2$ to the real mass squared.
Anyway, the integration over $\bfq$ in Eq.~(\ref{varphi-Conjvarphi}) 
can be done by the usual formula for the Dirac delta function as
\begin{equation}
f^2
\int\frac{4\pi q^2dq}{(2\pi)^3}
\frac{\pi}{2\omega_q\,\omega^*_q}\,
\left[ 2\Re\Big(\frac{d\omega_q}{dq}\Big)\right]_{q=q_{\rm CM}(E)}^{-1} 
\delta(q- q_{\rm CM}(E) )
=\frac{f^2}{4\pi}\,\frac{q_{\rm CM}(E)}{2E}.
\label{varphi-Cvarphi2}
\end{equation}
 
Next consider the `ghost-ghost' and  
`anti-ghost--anti-ghost' production whose probability 
is given as follows by again writing $P^0=2E$ and using the property 
$\delta_\rmc(2z)=(1/2)\delta_\rmc(z)$:
\begin{align}
\frac{f^2}{8}
\int\frac{d^3\bfq}{(2\pi)^3}
\Bigl(\frac{\pi}{\omega_\bfq^2}\,
\delta_\rmc(E-\omega_\bfq) 
+\frac{\pi}{{\omega^*_\bfq}^2}\,
\delta_\rmc(E-\omega^*_\bfq)
\Bigr) \,.
\label{GhostGhost}
\end{align} 
In subsection \ref{sec:complex-delta}, we have considered some properties of this complex delta function 
$\delta_\rmc(E-\omega_\bfq)$ as a function of $E$ with a fixed complex $\omega_\bfq$.
Here, we have to integrate this function over the variable $q=|\bfq|$ 
and need its property as a function of $q$, though we have no convenient 
formula for the complex delta function as we do for Dirac's delta, i.e.,
\begin{equation}
\delta\big(f(x)\big) = \sum_{x_0^{(i)}\text{:\ zeros of\ } f(x)} 
\left|\frac{df}{dx}(x_0^{(i)})\right|^{-1} \delta\big(x-x_0^{(i)}\big)  \,.
\label{DiracDeltaFml}
\end{equation}
This is the formula for the real function $f(x)$ of real variable $x$, which  
we have just used above. We shall show a similar formula holds 
for some cases of the present complex delta function $\delta_\rmc(E-\omega_\bfq)$.
In order to treat this problem properly, we use the adiabatic (Gaussian) 
regularization form of the complex delta function, so we discuss
\begin{align}
I(E) = \lim_{a \rightarrow0} 
\int\frac{q^2 dq}{\omega_q^2}\,\Delta_a(E-\omega_q) \,,
\label{I(E)}
\end{align} 
where we recall that
\begin{align}
\Delta_a(E-\omega_q) &= \frac1{2a\sqrt\pi} e^{-G(q)/4a^2}  
= \frac1{2a\sqrt\pi} \exp\left[-\frac{G_R(q)}{4a^2}\right]\cdot e^{-iG_I(q)/4 a^2}\,,
\nn
G(q)&= (E-\omega_q)^2, \qquad \omega_q = \sqrt{q^2 + m^2+i\gamma^2}\,,\nn
G_R(q)&= \Re G(q) = 
(E -\Re\omega_q+\Im\omega_q)(E -\Re\omega_q-\Im\omega_q)\,, \nn
G_I(q)&= \Im G(q) =- 2(E-\Re\omega_q)\Im\omega_q \,.
\label{eq:Gs}
\end{align}
Here, the real part of $I(E)$ multiplying the const., i.e., \ $(f^2/16\pi)\times2\Re[I(E)]$, gives the above 
`ghost-ghost' and  `anti-ghost--anti-ghost' production probability 
(\ref{GhostGhost}).

\begin{figure}[ht]
\begin{center}
\hspace{0.7cm}
\includegraphics[width=8cm]{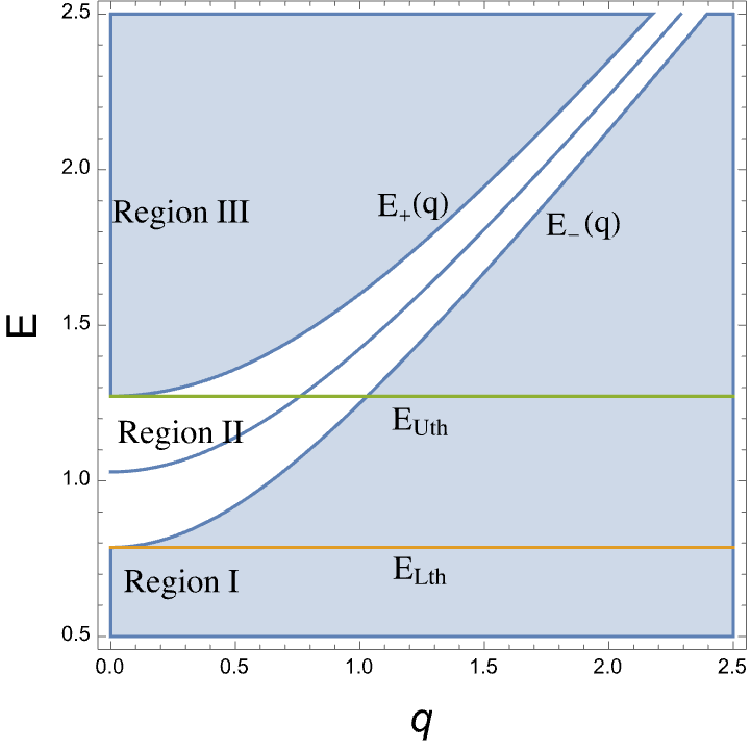}
\caption{The unshaded region is the negative $G_R(q)$ region, i.e., 
where $G_R(q)<0$, and thus gives the support of the complex delta 
function $\delta_\rmc(E-\omega_q)$. 
The upper and lower boundary curves of the negative $G_R(q)$ region 
are given by $E = \Re\omega_q\pm\Im\omega_q\equiv E_{\pm}(q)$. The curve in the 
middle of the region is $E=\Re\omega_q$ drawn for reference. 
This figure is drawn with parameter $\gamma^2/m^2=0.5$. 
The lowest horizontal line (orange) is at lower threshold energy 
$E=E_-(q{=}0)\equiv E_\text{Lth}=0.786$ and the upper horizontal line (green) is at 
the upper threshold energy $E=E_+(q{=}0)\equiv E_\text{Uth}=1.272$.
$G_R(q)=0$ has no  (region I), one (region II) and two solutions
(region III) of real $q$.}
\label{negativeGR}
\end{center}
\end{figure}

As noted in Sect.~3, the real part $G_R(q)$ of the exponent factor 
$G(q)=(E-\omega_q)^2$ determines the magnitude of the regularized complex delta 
function $\Delta_a(E-\omega_q)$. In particular, 
in the $a\rightarrow0$ limit, its sign is crucial since
the form $\propto\exp\big(- G_R(q)/ 4a^2\big)$ implies that 
when $G_R(q)>0$ the limit vanishes while it diverges (and rapidly oscillates) 
if $G_R(q)<0$. 
That is, the complex delta function (distribution)
$\delta_\rmc(E-\omega_q)$ defined as $\lim_{a\rightarrow0} \Delta_a(E-\omega_q)$, 
when viewed as the function of $E$ for each fixed $q$, 
has non-vanishing support only in the {\it negative} $G_R(q)$ region. 
Since $G_R(q)$ is quadratic in $E$, the negative $G_R(q)$ region is 
simply given by  
\begin{equation}
E_-(q) < E < E_+(q),   \qquad E_{\pm}(q) = \Re\omega_q\pm\Im\omega_q   \,.    
\end{equation}
We can plot the positive/negative $G_R(q)$ region, or the lower boundary 
curve $E=E_-(q)$ and also the upper boundary curve $E= E_+(q)$ as a function 
of $q$ in the real $(q, E)$ plane. 
See Fig.~\ref{negativeGR} in which these functions are plotted
for the choice of parameter $\gamma^2/m^2=0.5$.  
As it is easily shown, $\Re\omega_q$ ($\Im\omega_q$) is a monotonically increasing 
(decreasing) function of $q(>0)$ and both curves $E=E_{\pm}(q)$ are 
monotonically increasing.
We have called the boundary energies $E_-(0)$ and 
$E_+(0)$ at $q=0$ the {\it lower threshold} $E_\text{Lth}$
and the {\it upper threshold} $E_\text{Uth}$, respectively, in 
Sect.~3. They are the lowest points of the lower and upper 
boundary curves $E=E_{\mp}(q)$, respectively.     

Drawing horizontal lines in this figure on the plane $(q,\,E)$ 
at each constant energy $E$, 
the following facts for the three energy regions become immediately 
clear from Fig. \ref{negativeGR}:
\begin{enumerate}
\item Region I : Energy region below the lower threshold, $E\le  E_\text{Lth}$. 

Each horizontal line, from $q=0+$ to $\infty$, at energy $E \le E_\text{Lth}$, 
runs solely in the shaded region, i.e., in the positive $G_R(q)>0$ region, 
so that $\delta_\rmc(E-\omega_q)$  vanishes for $\forall q>0$. 

\item Region II : Energy region in between the lower and upper thresholds,
$E_\text{Lth} < E \le  E_\text{Uth}$.

Each horizontal line at an energy $E$ in between 
$E_\text{Lth}< E \le  E_\text{Uth}$  
starts at $q=0+$ in the negative $G_R(q)<0$ region and  crosses the 
lower boundary curve at a point $q=q_+(E)$ satisfying $E=E_-\big(q_+(E)\big)$, 
beyond which the line runs into the positive $G_R(q)>0$ region. 
This implies that the support of the complex delta function in 
this energy region II is the interval $0<q < q_+(E)$. 

\item Region III : Energy region above upper threshold, $E > E_\text{Uth}$.

Each horizontal line at an energy $E$ above the upper threshold $E > E_\text{Uth}$ 
starts at $q=0$ in the positive $G_R(q)>0$ region and  crosses the 
upper boundary curve at a point $q=q_-(E)$ satisfying 
$E=E_+\big(q_-(E)\big)$ and enters the negative $G_R(q)<0$ region.   
It then crosses the lower boundary curve at a point $q=q_+(E)$ 
satisfying $E=E_-\big(q_+(E)\big)$, 
beyond which the line again enters into the positive $G_R(q)>0$ region. 
This implies that the support of the complex delta function 
in this energy region III 
is the interval $ q_-(E) <q < q_+(E)$. The momentum $q$ in the middle  
in this interval becomes larger as $E$ becomes larger, but the interval 
width becomes smaller.   
\end{enumerate}

We have already understood the $q$-integration of (\ref{I(E)}) for the energy $E$
below the lower threshold. It vanishes since $\delta_\rmc(E-\omega_q)$ vanishes 
for $\forall q\in[0,\infty]$ in the energy Region I. The 
production of ghost-ghost pair does not occur, justifying the name of 
`lower threshold energy' for $E_\text{Lth}=\Re\omega_0 -\Im\omega_0$.

Now the non-trivial problem is how one may evaluate the  $q$-integration 
of (\ref{I(E)}) for the energies $E$ in Regions II and III. 
At this stage, we recall our basic strategy to evaluate the integral of 
the complex delta function $\delta_\rmc$. That is the change of the integration 
variable by making the contour shift or deformation in such a way as to 
make the argument of $\delta_\rmc$ real. Then, $\delta_\rmc$ is reduced to the 
usual Dirac delta function $\delta$ to which we can apply the formulas like 
Eq.~(\ref{DiracDeltaFml}). 

It is in fact an easy task to make the argument of our present 
delta function $\delta_\rmc(E-\omega_q)$ real by shifting the integration variable 
$q$ in the complex $q$ plane. The imaginary part of the argument $E-\omega_q$ 
comes solely from 
\begin{equation}
\omega_q = \sqrt{ q^2 + m^2 + i\gamma^2}.
\end{equation}
If we change the variable $q^2=:z$, which originally runs from 0 to $+\infty$ 
in the integral (\ref{I(E)}), into 
\begin{equation}
 z = x - i\gamma^2 \,, 
 \label{eq:z}
\end{equation}  
then $\omega_q$ becomes $\sqrt{ x + m^2 }$ (like a particle energy of real mass $m$),
which remains real as far as $x$ runs over real positive axis $[0,\,\infty]= R_+$. 
We call this line of $z$ swept by $x\in R_+$  ``contour $R_+(q_0)$" 
for a reason which becomes clear shortly. On the complex $z=q^2$ plane, 
$R_+(q_0)$ is a straight line parallel to the real axis (though on the 
complex $q$ plane, it looks like a part of a square root curve of course). 
So, for the evaluation of the original integral in Eq.~(\ref{I(E)}), 
we should make the following contour deformation on the complex $z$-plane:
\begin{align}
R_+[ 0 \ \rightarrow\ \infty] \ \ \Rightarrow\quad 
&C_1[ 0+ i0 \ \rightarrow\ 0 - i\gamma^2 ]  \nn
&+R_+(q_0)[ 0 - i\gamma^2 \ \rightarrow\ +\infty- i\gamma^2 ]  \nn
&+C_2[ +\infty- i\gamma^2 \ \rightarrow\ +\infty+ i0 ].  
\label{Deform2Rq0}
\end{align}
We can forget the contribution from the last vertical line $C_2$ 
at $\Re z=\infty$ which 
vanishes as usual. 
And, as we will do shortly, the contribution from the horizontal contour 
segment $R_+(q^0)$ can be 
very easily evaluated since $E-\omega_q$ is real on the whole contour $R_+(q^0)$
and hence, the complex delta function $\delta_\rmc(E-\omega_q)$  is reduced 
to the usual Dirac delta function $\delta(E-\omega_q)$ of real variable $x$.

Before doing this task, let us examine the contribution from the 
first vertical line $C_1$ in which a critical difference appears between 
the cases with energy in Region II and Region III.

\vskip 1ex

\noindent
\underline{Region} II: $E_\text{Lth} <  E \le  E_\text{Uth}$ 

\vskip 1ex

As we see in Fig.~\ref{negativeGR}, 
the $q=0+$ point in Region II is in the negative $G_R(q)$ 
(unshaded) region. This must also be the case even when 
$q$ is extended to be complex; around the 
origin $q=0$ in the complex $q$ plane, the real part of $G(q)$, $G_R(q)$, 
is  negative.     
This means that the complex delta function 
$\delta_\rmc(E-\omega_q)=\lim_{a\rightarrow0}\Delta_a(E-\omega_q)$ becomes divergent 
(and rapidly oscillating) in the $a\rightarrow0$ limit 
in a finite neighborhood of $q=0$.\footnote{%
This divergence of the complex delta function 
$\delta_\rmc(E-\omega_q)$ as a function of $q$ for the energy $E$ in Region II 
is essentially the same as that of $\delta_\rmc(E-\omega_\bfP)$ for the single 
ghost production in Sect. 3.} 

So, the $a\rightarrow0$ limit of the integral (\ref{I(E)}) is already divergent 
in the contribution from the contour segment $C_1$ alone. 
Therefore, there is no particular reason to adopt the deformed contour 
(\ref{Deform2Rq0}). We use the original integration contour 
$R_+$. 

\def\barE{\overset{\,\rule[-.5pt]{.5em}{.5pt}}{E}}   
\def\subbarE{\overset{\,\rule[-.5pt]{.4em}{.5pt}}{E}}  

Although it is divergent, the integral $I(E)$ has a well-defined meaning 
as a distribution. We average the integral (\ref{I(E)}) by using 
the Gaussian smearing function 
(\ref{GaussianSmear}) around an energy $\barE$ in Region II with 
standard deviation $\sigma$ leading to
\begin{align}
I(\barE) &= \int_{-\infty}^\infty dE\, f_{\subbarE}(E)\, I(E) \nn
&=
\int_0^\infty dq\,\frac{q^2}{\omega_q^2}\,
\frac1{\sqrt{2\pi}\sigma}\exp 
\left[-\frac12\Bigl(\frac{\omega_q - \barE}{\sigma}\Bigr)^2\right] \,.
\end{align}
This now gives a well-defined finite value thanks to 
the finite width of $\sigma$ although there is some interval $0 \le q<  q_+$ 
where $\Re[(\omega_q - \bar E)^2]<0$ for the present energy $\bar E$  in Region II.

\vskip 1ex

\noindent
\underline{Region} III: $E > E_\text{Uth}$ 

\vskip 1ex

Contrary to the above energy Region II, the Fig.~\ref{negativeGR}  
shows that the $q=0$ point  in Region III is in the positive 
$G_R(q)$ (shaded) region. Again, by analyticity, this must 
be the case 
around $q=0$ also in the complex $q$-plane or complex $q^2= z$ plane. 
Actually, for energy $E$ in this Region III, we can easily show that the 
whole vertical line $C_1$ is in the positive $G_R(q)$ region as follows.
 
On the vertical line $C_1$, the complex variable $z$ is parametrized as 
$z = 0 -iy$ by the real parameter $y\in[0,\,\gamma^2]$ and $\omega_q$ looks like 
$\omega_q=\sqrt{m^2 + i(\gamma^2-y)}$. As we go down from the origin $z=0$ to the 
point $z=-i\gamma^2$ touching $R_+(q_0)$, the real parameter $y$ changes from 
0 to $\gamma^2$ and the real part $G_R(q)$ of $G(q)=(E-\omega_q)^2$ monotonically 
{\em increases}:
\begin{align}
&\frac{dG(q)}{dy} = 2(E-\omega_q)\frac{i}{2\omega_q} =
i\,\frac{E}{\omega_q}-i \quad \rightarrow\quad  \nn
&\frac{dG_R(q)}{dy} = 
\Re\Big[\frac{dG(q)}{dy}\Big] = 
\Re\Big[i\,\frac{E}{\omega_q}-i\Big] = 
-E\,\Im\Big[\frac1{\omega_q}\Big] = 
\frac{E}{|\omega_q|^2}\,\Im\omega_q \geq0 \,.  
\end{align}
Since it is positive at the starting point $z=0$ for 
the energy $E$ in the Region III, the $G_R(q)$ keeps positive 
all the way along $C_1$. 

The positivity of $G_R(q)$ on $C_1$ means that the 
contribution from the segment $C_1$ to the integral (\ref{I(E)}) vanishes 
in the $a\rightarrow0$ limit. 
Thus the contribution can come only from the horizontal line $R_+(q_0)$. 

Now, $\omega_q$ is real on $R_+(q_0)$ as stated above,  
so $G(q)=(E-\omega_q)^2$ is positive aside from the possible isolated 
zeros. The zeros, say $q=q_0$, of $G(q)$ are also easily found: 
\begin{align}
&G_(q_0)=0 \ \ \rightarrow\quad \omega_{q_0}=E 
\ \ \rightarrow\quad  q_0^2 = E^2 - m^2 -i\gamma^2 \nn
& \hbox{i.e.,} \ \ \rightarrow\quad 
\Re(q_0^2) = E^2- m^2, \quad \ \  \Im(q_0^2)= -\gamma^2\,.
\end{align}  
That is, aside from the sign $\pm$, the zero $q_0$ is uniquely given by 
\begin{equation}
q_0 = \big(E^2 - m^2 -i\gamma^2\big)^{1/2} 
= \sqrt{E^2 - m^2}\Big( 1 -i\frac{\gamma^2}{E^2-m^2}\Big)^{1/2} \,.
\end{equation} 
We denote the zero with $\Re q_0>0$ as $q_0$ and the other as $-q_0$. 
The one on $R_+(q_0)$ is $q_0$, so the name $R_+(q_0)$ means that it passes 
through the point $q_0$.
We now know that the complex delta function $\delta_\rmc(E-\omega_q)$ becomes the 
usual Dirac delta function if written in terms of the real variable 
$x= z +i\gamma^2= q^2+i\gamma^2$ on $R_+(q_0)$ and it has only one zero 
$x_0 = q_0^2 + i\gamma^2= E^2-m^2$. 
We can apply the formula (\ref{DiracDeltaFml}) to $\delta_\rmc(E-\omega_q) 
= \delta\big(f(x)\big)$ with $f(x)=E-\omega_q$ to evaluate  
the contribution of this contour segment $R_+(q_0)$ 
to the integral (\ref{I(E)}).
\begin{align}
&\frac{df}{dx}= -\frac{d\omega_q}{dq}\, \frac{dq}{dx} 
= -\frac{2q}{2\omega_q}\,\frac1{2q}=-\frac1{2\omega_q}, \qquad 2q\,dq = dx, \nn 
&\delta\big(f(x)\big) = (2\omega_q)\Big|_{q=q_0}\cdot \delta(x-x_0) = 2E\,\delta(x-x_0),\nn
&
I(E)  
= \int_{R_+(q_0)} dq\,\frac {q^2}{\omega_q^2}\delta_\rmc(E-\omega_q)
= 
\int_{R_+(q_0)} dx\,\frac {q}{\omega_q}\delta(x-x_0) 
=\frac {q_0}{\omega_{q_0}}= \frac {q_0}{E}\,.
\label{eq:IEfinal}
\end{align}
It is truly remarkable that such a non-trivial integral $I(E)$ of the 
complicate function $(q^2/\omega_q^2)\delta_\rmc(E-\omega_q)$ of $q$ 
including complex delta 
function, which is actually a distribution possessing non-localized support, 
can be evaluated analytically by computing a single complex root $q_0$ of 
$G(q)=0$. 

However, to verify the equivalence of the original integration over $R_+$  
with that over the deformed one (\ref{Deform2Rq0}), 
we must confirm that the function $q^2/\omega_q^2$ multiplied 
by the complex delta function $\delta_\rmc(E-\omega_q)$ in Eq.~(\ref{I(E)}) 
has no pole singularities in the 
rectangular domain surrounded by $R_+$ and 
$C_1+ R_+(q_0)+C_2$ in the $z$ plane. 
The only singularity is the pole at $\omega_q^2=0$, i.e.,\ at 
$z= -m^2-i\gamma^2$ which is just on the negative side line $R_-(q_0)$ 
of our integration contour segment $R_+(q_0)$, thus indicating clearly 
that the pole is outside the rectangular domain.   

\begin{figure}[ht]
\begin{center}
\hspace{0.7cm}
\includegraphics[width=10cm]{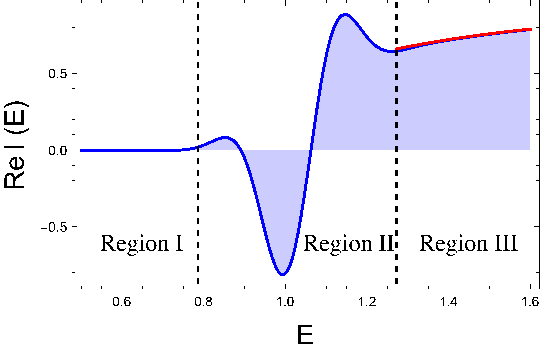}
\caption{Comparison of the numerical calculation (blue) of
$\Re I(E)$ with that of the exact result (red) 
(\ref{eq:IEfinal}), where
$m=1$ and $ \gamma=1/\sqrt{5}$ with a finite cutoff $a=1/14$
are used. The vertical lines separate the three energy regions, where
the threshold values are: $E_\text{Lth}=0.7862\,, E_\text{Uth}=1.272$.
}
\label{figIE}
\end{center}
\end{figure}

In Fig. \ref{figIE} we compare the exact result (\ref{eq:IEfinal}) with
a numerical calculation, where we have used:
$m=1\,, \gamma=1/\sqrt{5}$ with a finite cutoff $a=1/14$.
The blue line presents the numerical
result of the real part of (\ref{I(E)}), and the red  line shows
the real part of  the exact result  (\ref{eq:IEfinal}),  which is applicable  
starting at $E=E_\text{Uth}=1.272$.
We see an excellent agreement of the numerical result with
the exact result already at finite $a=1/14$.

\subsubsection{Photon and ghost production}  
\label{sec:photon-ghost}

The production probability  of
photon and ghost, $ A$-$\varphi$ and $A$-$\varphi^\dag$, is given by 
the second line of (\ref{Norm-2phi}):
\begin{align}
 -f^2
\int\frac{d^3 \bfq}{(2\pi)^3}
&
\Bigl(\,\frac{\pi}{2\nu_\bfq\,\omega_\bfq}\,
\delta_\rmc(\nu_\bfq+\omega_\bfq-P^0) 
+\frac{\pi}{2\nu_\bfq\,\omega^*_\bfq}\,
\delta_\rmc(\nu_\bfq+\omega^*_\bfq-P^0)\,
\Bigr)\,.
\label{Norm-2phi-1}
\end{align} 
We therefore consider the integral 
\begin{equation}
I_\text{ph}(E) =\lim_{a\to 0} \int \frac{q^2 dq }{\nu_q\omega_q }\,
\Delta_a (E-(\nu_q+\omega_q)/2)\,,
\label{eq:IET}
\end{equation}
where 
\begin{align}
& \Delta_a(E-(\nu_q+\omega_q)/2) = \frac1{2a\sqrt\pi} e^{-\tilde{G}(q)/4a^2} \,, 
\nn
& \tilde{G}(q)= (E-(\nu_q+\omega_q)/2)^2\,, \,
\nu_q = \sqrt{q^2 + \delta^2}\,, \omega_q = \sqrt{q^2 + m^2+i\gamma^2}\,, 
\nn
&\tilde{G}_R(q)= \Re \tilde{G}(q) = 
\big(E -(\nu_q+\Re\omega_q+\Im\omega_q)/2)
(E -(\nu_q+\Re\omega_q-\Im\omega_q)/2\big) \,, 
\nn
&\tilde{G}_I(q)= \Im \tilde{G}(q) =- \big
(E-(\nu_q+\Re\omega_q)/2\big)\Im\omega_q \,.
\label{eq:Gtildes}
\end{align}
Comparing the expressions in (\ref{eq:Gtildes}) with those of (\ref{eq:Gs})
we can easily convince ourselves  that the same classification of the energy regions
exists here as in the case of two ghosts production:
The lower and upper threshold energies are given by
$E_\text{Lth}=(\nu_0+\Re \omega_0-\Im \omega_0)/2$ and
$E_\text{Uth}=(\nu_0+\Re \omega_0+\Im \omega_0)/2$, respectively.
In Region I ($E\le E_\text{Lth}$) 
the real part $\tilde{G}_R(q) $ is positive for $\forall q>0$, so that the integral
$I_\text{ph}(E)$ given in Eq. (\ref{eq:IET}) vanishes in the $a\to 0$ limit.
In Region II the integral diverges, which means  that we have to average it with the Gaussian smearing function 
(\ref{GaussianSmear}) to obtain a finite result.

\begin{figure}[ht]
\begin{center}
\hspace{0.7cm}
\includegraphics[width=10cm]{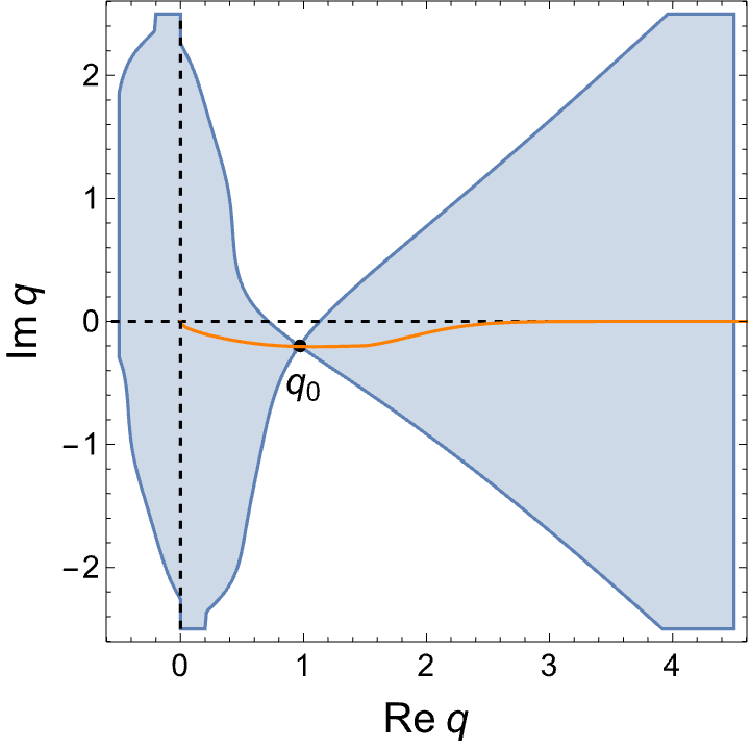}

\caption{The shaded area is the positive   ${\rm Re} \,\tilde{G}(q)$
region  on the complex plane of $q$.
The black point is $q_0$
(zero of $(\nu_q+\omega_q)/2-E$).
The orange line (contour) starts from the origin, goes through
$q_0$ and approaches $+\infty$, without leaving
the shaded area.
}
\label{positiveG}
\end{center}
\end{figure}

\begin{figure}[ht]
\begin{center}
\hspace{0.7cm}
\includegraphics[width=10cm]{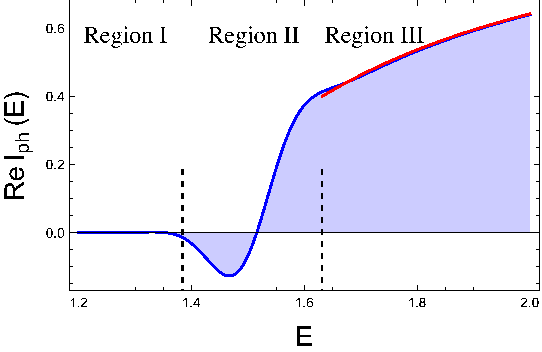}
\caption{Comparison of the numerical calculation (blue) of the real part of 
$I_\text{ph}(E)$ with that of the exact result (red) 
(\ref{eq:IETfinal}), where
$m=2$ and $ \gamma=1$ with a finite cutoff $a=1/20$
are used. The vertical lines separate the three energy regions, where
the threshold values are: $E_\text{Lth}=1.384\,, E_\text{Uth}=1.632$.
}
\label{figIET}
\end{center}
\end{figure}

The situation in Region III is slightly different:
There no longer exits such a simple variable like $z=q^2$ 
in Eq. (\ref{eq:z}) that the  line,  realizing real $\nu_q+\omega_q$
in the complex plane of that variable,  becomes a straight line
parallel to the real axis.
Fortunately, it is {\it not}  a necessary condition for 
the integral (\ref{eq:IET}) to be analytically evaluated by computing
a single complex root $q_0$:
It is sufficient for this  if there exists a curve (contour) 
on the complex $q$ plane, which starts from the origin
(i.e., $q=0$), goes through the zero $q_0$ of $(\nu_q+\omega_q)/2-E$ and
approaches $+\infty$, such that the real part of $\tilde{G}(q)$  is all the way
positive,
except at $q=q_0$. An example  is shown in Fig. \ref{positiveG}, where we have used representative values $E/m=0.9\,,
\gamma/m=0.5\,,\delta/m=0.5$ with
$m=2$.
As far as such a contour exists, the calculation  of the integral reduces 
to the evaluation of  the contribution  only in the infinitesimal neighbourhood of $q_0$.
We thus expand the argument of the complex delta function
around $q=q_0$:
\begin{eqnarray}
& &(\nu_q+\omega_q)/2-E =
(q-q_0)A+O((q-q_0)^2)\,,\\
& &\mbox{where} ~~
\left.\big((\nu_q+\omega_q)/2-E\big)\right|_{q=q_0}=0~\mbox{and}~
A =\left.\frac{d}{d q} \big((\nu_q+\omega_q)/2-E\big) \right|_{q=q_0}\,.
\nonumber
\end{eqnarray}
Therefore,  $G(q)$ near $q=q_0$ can be written as
\begin{equation}
\tilde{G}(q) =(q-q_0)^2\,A^2+\cdots\,,
\end{equation}
which means that integral (\ref{eq:IET}) becomes a Gaussian integral
(with a complex coefficient $A$). 
In this way we arrive at
\begin{equation}
I_\text{ph}(E)=\left.A^{-1}\frac{q^2}{\nu_q \omega_q}\right|_{q=q_0}
=\frac{q_0}{E}\,.
\label{eq:IETfinal}
\end{equation}
In Fig. \ref{figIET} we compare the exact result (\ref{eq:IETfinal}) with
a numerical calculation, where we have used
$m=2\,, \gamma=1$ with a finite cutoff $a=1/20$.
The blue line presents the numerical
result of the real part of (\ref{eq:IET}) and the red line 
shows the real part of the analytic result  (\ref{eq:IETfinal}), which is applicable  
above the energy 
$E=E_\text{Uth}=1.632$.
We thus also see here an excellent agreement of the numerical result with
the exact result.
 
 \section{Conclusion}
 \label{sec:conclusion}
In this paper, we have examined whether complex ghosts are truly 
not created by collisions of (positive norm) physical particles, 
as claimed by Lee and Wick \cite{Lee:1969fy,Lee:1969zze,Lee:1970iw}. 
This problem is 
of very general interest because, if their claim is 
true, all theories can in principle be made renormalizable or even finite 
without violating unitarity, simply by adding higher derivative 
regulators. 
More importantly, quadratic gravity theory which comes with a natural   
built-in regulator, becomes a viable perturbatively renormalizable 
theory of gravity. 
To the author's knowledge, no clear disproof nor sound proof 
has been given for physical unitarity itself, aside from some severe  
criticisms on the Lorentz invariance \cite{Nakanishi:1971jj,Gleeson:1971cvx} and causality \cite{Lee:1969fy,Coleman:1969xz,Donoghue:2019fcb,Donoghue:2021eto}. 

Our answer to this question is: Complex ghosts can in fact be created 
with finite (non-zero) probability by the collisions 
of physical particles alone, thus implying that physical unitarity is violated. 
This is a clear and inevitable conclusion 
as far as one works faithfully within the QFT framework. 
It should also be emphasized that there is no room 
 for inequivalent changes of  the integration contours in the complex plane of the 
energy-momentum variables in Feynman diagrams.  In other words, if one finds 
a clever method to change the integrations in such a way as to satisfy unitarity,
 it is no longer a quantum field theory. Furthermore, if one leaves the operator 
formalism of QFT, it would be very difficult to prove the unitarity 
in closed form. 

We were able to avoid the cumbersome (and sometimes difficult) considerations 
of the integration contours of the energy variables by using the $3d$ momentum 
form  for the propagators, which amounts to a great 
simplification attained at the sacrifice of apparent covariance. 
We note, however,  that it may be desirable to develop a more systematic computation technique 
by using the $4d$ expression propagators. 

Although the energy conservation becomes slightly obscured by the 
appearance of a complex mass squared, we have found a clear 
threshold energy for the creation of the complex ghost, that is, the 
lower threshold given by $E_{\rm Lth}=\Re\sqrt{M^2}-\Im\sqrt{M^2}$. 
At energies below this threshold, there occurs no ghost production, meaning that unitarity and renormalizability hold. Consequently, we have
 a consistent effective QFT for  $E<E_{\rm Lth}$.

The complex delta function has played a central role in this complex ghost 
theory, and its property as a distribution is very natural and mathematically 
consistent. Moreover, it can be used whenever states with complex energy 
eigenvalue appear, irrespective of the norm. Indeed, 
Nakanishi \cite{Nakanishi:1958,Nakanishi:1958-2} introduced 
this complex delta function to construct a complex energy eigenstate 
corresponding to the usual (i.e., positive norm) unstable particle. 
It may thus be interesting to develop a QFT of unstable particles and 
compare it with the present complex ghost theory in order to answer the question: What is the essential 
difference between an unstable particle and a complex ghost?





\section*{Acknowledgment}
We thank Takashi Ichinose, Jeffrey Kuntz and  Noboru Nakanishi
for valuable comments and suggestions.
We also thank
Manfred Lindner, Jonas Rezacek, Philipp Saake,
Andreas Trautner and Masatoshi Yamada for useful discussions.
We further  thank Bob Holdom for critical discussions 
on asymptotic complex ghost states.
This work was supported in part by the MEXT/JSPS KAKENHI Grant Number
JP18K03659 (T.K.) and 23K03383 (J.K.).

\appendix

\section{Evaluation of the forward scattering amplitude (\ref{eq:S-1-6})}

In this Appendix we evaluate the amplitude Eq.~(\ref{eq:S-1-6}) in terms of 
4d form propagators explicitly in the CM frame ($\bfP =0$) and 
confirm that it reproduces the result (\ref{eq:S-1-5}) obtained from the
3d form of propagators. 
Note that since $\bfq=-\bfq'$ in the CM frame, 
we have $\omega_\bfq =\omega_{\bfq'}$.
But we keep the notation $\omega_{\bfq'}$ in order to trace
where the pole comes from.

We first rewrite the $q'^0$-integration 
for the ghost propagator in Eq.~(\ref{eq:S-1-6}) 
along the curved contour $C$ in Fig. \ref{figC} (left) to 
the contour shown in Fig. \ref{figC} (right) as was done in (\ref{deform2C2}); 
in this case, the propagator is multiplied by $\delta_\rmc(P^0-q^0-q'^0)$ 
which may be regarded as an analytic distribution of the variable $q'^0$ 
as explained in the footnote to Eq.~(\ref{ghost-cont}), so we can 
perform this deformation of the $q'^0$ integration contour and find 
\begin{equation}
\int_C dq'^0 \frac1{\omega_{\bfq'}^2-{q'^0}^2} \delta_\rmc(P^0-q^0-q'^0)= 
\int_R dq'^0 \frac1{\omega_{\bfq'}^2-{q'^0}^2} \delta_\rmc(P^0-q^0-q'^0) + 
\frac{2\pi i}{2\omega_{\bfq'}} \sum_{\pm}\delta_\rmc(P^0-q^0\pm\omega_{\bfq'})\,, 
\label{ghostFormula}
\end{equation}
where the last terms are contributions from the two poles of 
the ghost propagator at $q'^0=\mp \omega_\bfq$.
The first term $\int_R$ is now the $\varphi$ propagator given by 
the $q'^0$ integral along the real axis $R$ in the same way as the
`photon' $A$ and anti-ghost $\varphi^\dagger$.

Applying this formula (\ref{ghostFormula}) to the $\varphi$ ghost propagator 
parts in Eq.~(\ref{eq:S-1-6}), we find
\begin{align}
&\bra{I(\bfp_1, \bfp_2)}\hat T\ket{I(\bfp_1, \bfp_2)}\big|_{\text{$s$-ch.}} 
\nn
&=f^2 \int\frac{d^3\bfq}{(2\pi)^3}
\biggl\{
\sum_{a,b}\eta_a\eta_b\int_R \frac{dq^0}{2\pi i} \int_R\,dq'^0\, 
\frac{1}{{\omega^a_\bfq}^2-{q^0}^2}\,
\frac{1}{{\omega^b_{\bfq'}}^2-{q'^0}^2}\,\delta_\rmc(P^0-q^0-q'^0) \nn
&\hspace{6.8em}{} +2\sum_{a} \eta_a \eta_\varphi 
\int_R dq^0 \frac1{{\omega^a_\bfq}^2-{q^0}^2}\,
\cdot \sum_{\pm}\frac1{2\omega_{\bfq'}}
\delta_\rmc(P^0-q^0\pm\omega_{\bfq'}) \nn
&\hspace{6.8em}{} + (\eta_\varphi)^2 \sum_{\pm,\pm} 
\frac{2\pi i}{4\omega_\bfq\omega_{\bfq'}} \delta_\rmc(P^0\pm\omega_\bfq\pm\omega_{\bfq'})  
\biggr\}
,\label{eq:T-6}
\end{align}
where $\eta_\varphi=-1/2$.   
Here the first line is the contribution from all the terms where 
both $q^0$ and $q'^0$ integrations are along real axis $R$, so that the 
complex delta function $\delta_\rmc(P^0-q^0-q'^0)$  reduces to the Dirac one 
and the $dq'^0$ integration becomes trivial to set $q'^0$ equal to $P^0-q^0$. The $dq^0$ integration becomes 
non-trivial however, as we will see 
shortly. 
The second line comes from the cross terms of the $\phi_a$ propagator 
(integrated on $R$) and the last $\delta_\rmc$ factor in the formula 
(\ref{ghostFormula}).     
The third line comes from the term in which both $\phi_a$ and $\phi_b$ are 
ghosts $\varphi$, and both the $dq^0$ and $dq'^0$ integrations pick up 
the ghost poles. \\

[I] {\it The $dq^0$ integral of the first line} (after the trivial $dq'^0$ integration) 
can be done by closing the integration contour $R=(-\infty,\,\infty)$ after adding a
half circle at infinity either in the lower or upper half complex $q^0$-plane
and applying Cauchy's theorem. 
Both ways of choosing the contour give the same result, so we choose the contour 
closing in the lower-half plane so that the poles in the lower-half plane 
are picked up. The poles existing in the lower-half plane are of positive 
frequency, $+\nu_\bfq$ for the `photon' $A$ and $+\omega^*_\bfq$ 
for anti-ghost $\varphi^\dagger$, 
while it is of negative frequency $-\omega_\bfq$ for the ghost. 
Denoting them collectively as $(+\nu_\bfq,\,-\omega_\bfq,\,+\omega^*_\bfq)=: 
\omega^{La}_\bfq $, then noting ${\omega^a_\bfq}^2={\omega^{La}_\bfq}^2$ and also 
that the poles of the $q'$ propagator 
(of field $\phi_b$) 
in the lower-half plane are placed at $q^0=P^0+\omega^{Lb}_{\bfq'}$, 
we have the following for the first line 
integral
\begin{align}
&\int_R \frac{dq^0}{2\pi i} \int dq'^0\,
\frac1{{\omega^a_\bfq}^2-{q^0}^2}\,
\frac1{{\omega^b_{\bfq'}}^2-{q'^0}^2}\,\delta_\rmc(P^0-q^0-q'^0) \nn
&\quad{}=
\int_R \frac{dq^0}{2\pi i} 
\frac1{{\omega^a_\bfq}^2-{q^0}^2}\,
\frac1{{\omega^b_{\bfq'}}^2-(P^0-q^0)^2}  \nn 
&\quad{}=
\frac1{2\omega^{La}_\bfq}\,
\frac1{{\omega^{Lb}_{\bfq'}}^2-(P^0-\omega^{La}_\bfq)^2}  
+\frac1{{\omega^{La}_\bfq}^2-(P^0+{\omega^{Lb}_{\bfq'}})^2}\,
\frac1{2\omega^{Lb}_{\bfq'}} \ .
\end{align}
This last expression can be rewritten into a simpler form by 
using the decomposition into the 
partial fractions and a cancellation among them as follows:
\begin{align}
&= \frac1{2\omega^{La}_\bfq\, 2\omega^{Lb}_{\bfq'}} \left(\biggl(
\frac1{\omega^{Lb}_{\bfq'}-(P^0-\omega^{La}_\bfq)}+
\frac1{\omega^{Lb}_{\bfq'}+(P^0-\omega^{La}_\bfq)}\biggr)\right. \nn
&\hspace{10em}{}+\left.\biggl(\frac1{\omega^{La}_{\bfq}-(P^0+\omega^{Lb}_{\bfq'})}+
\frac1{\omega^{La}_{\bfq}+(P^0+\omega^{Lb}_{\bfq'})}
\bigg)\right) \nn
&= \frac1{2\omega^{La}_\bfq\, 2\omega^{Lb}_{\bfq'}} \left(
\frac{2(\omega^{Lb}_{\bfq'}+\omega^{La}_\bfq)}
{(\omega^{Lb}_{\bfq'}+\omega^{La}_\bfq)^2-{P^0}^2} \,
\right)\,.
\label{FirstLine}
\end{align}

[II] {\it Next we consider the second line integral in Eq.~(\ref{eq:T-6}). }
In order to make the argument of $\delta_{\rmc}(P^0-q^0+\omega_{\bfq'})$ 
(resp. $\delta_{\rmc}(P^0-q^0-\omega_{\bfq'})$) 
real, we lift (lower)  the original $q^0$-integration contour $R$ to the 
contour $R(+\omega_{\bfq'})$ ($R(-\omega_{\bfq'})$) which is parallel to $R$ 
and passes the point $+\omega_{\bfq'}$ ($-\omega_{\bfq'}$).   
While shifting the $q^0$ contour from $R$ to $R(\pm\omega_{\bfq'})$ 
we encounter the poles of the propagator of $\phi_a= A$, $\varphi^\dagger$ or 
$\varphi$, 
\begin{equation}
\begin{cases}
q^0 = \mp(\nu_{\bfq}-i\varepsilon)  &\text{for }\ \ \phi_a=A \\
q^0 = \mp(\omega^*_{\bfq}-i\varepsilon)  &\text{for }\ \ \phi_a=\varphi^\dagger\\
q^0 = \pm(\omega_{\bfq}+i\varepsilon)  &\text{for }\ \ \phi_a=\varphi
\end{cases},
\label{PoleAvoid}
\end{equation}
respectively. 
Here the `photon' pole position $\nu_\bfq-i\varepsilon$ is the one with the usual 
$-i\varepsilon$ prescription. However, we have also shifted the poles of the 
ghost and anti-ghost propagators with the $\pm i\varepsilon$ trick 
because the anti-ghost pole $q^0 = \mp \omega^*_{\bfq}$ and ghost pole 
$q^0= \pm\omega_\bfp$ are on the 
shifted contour $R(\pm\omega_{\bfq'})$, respectively, since $ \pm\Im\omega_{\bfq'}$
is the same as $\mp\Im \omega^*_\bfq$ and $\pm\Im\omega_\bfq$ 
in the CM frame $\bfP={\bf0}$, meaning that we have to 
specify how to avoid the pole singularity on the new integration contours
$R(\pm\omega_{\bfq'})$. 
It turns out that either shift, $+i\varepsilon$ or $-i\varepsilon$, of the pole position
leads to the same final result, as well see below. So, we use $-i\varepsilon$ shift for the anti-ghost 
and $+i\varepsilon$ shift for ghost field as shown in Eq.~(\ref{PoleAvoid}), 
since the calculations below become slightly simpler with this choice. 
Then, the shift of the integration contour 
from $R$ to $R(\pm\omega_{\bfq'})$  passes through the `photon' pole 
$\mp(\nu_\bfq-i\varepsilon)$ as far as $\Im M^2$ is positive and  finite:
It  reaches neither the anti-ghost pole $\mp(\omega^*_{\bfq}-i\varepsilon)$ 
nor the ghost pole $\pm(\omega_\bfq+i\varepsilon)$ by an infinitesimal amount $i\varepsilon$. 
Therefore, we have to take care of the pole contribution only for the 
`photon' case $\phi_a=A$ in the second line integral in Eq.~(\ref{eq:T-6}):
We can deform the contour $R$ for the `photon' propagator to
\begin{align}
\int_R dq^0 \ &\rightarrow\ \int_{R(+\omega_\bfq)} + \int_{C(-\nu_\bfq+i\varepsilon)} \ \ \text{or}\ \ 
\ \rightarrow\ \int_{R(-\omega_\bfq)} - \int_{C(+\nu_\bfq-i\varepsilon)} 
\label{photonPole}
\end{align}
for the terms containing $\delta_\rmc(P^0-q^0+\omega_{\bfq'})$  and  
$\delta_\rmc(P^0-q^0-\omega_{\bfq'})$, respectively.  

However, let us start with simpler anti-ghost case $\phi_a=\varphi^\dagger$, 
for which we can shift the integration contour without meeting pole 
singularities:
\begin{align}
&\int_R dq^0\, \frac{1}{{(\omega^*_\bfq}-i\varepsilon)^2-{q^0}^2}\,
\cdot \sum_{\pm}\frac1{2\omega_{\bfq'}}
\delta_\rmc(P^0-q^0\pm\omega_{\bfq'}) \nn
&=\int_{R(+\omega'_\bfq)} dq^0\, \frac{1}{{(\omega^*_\bfq}-i\varepsilon)^2-{q^0}^2}\,
\cdot \frac1{2\omega_{\bfq'}}
\delta_\rmc(P^0-q^0+\omega_{\bfq'}) \nn
&\qquad {}+\int_{R(-\omega'_\bfq)} dq^0\,  \frac{1}{{(\omega^*_\bfq}-i\varepsilon)^2-{q^0}^2}\,
\cdot \frac1{2\omega_{\bfq'}}
\delta_\rmc(P^0-q^0-\omega_{\bfq'}) \nn
&= 
\frac1{2\omega_{\bfq'}} 
\left(
\frac{1}{{(\omega^*_\bfq}-i\varepsilon)^2-(P^0+\omega_{\bfq'})^2}
+\frac1{{(\omega^*_\bfq}-i\varepsilon)^2-(P^0-\omega_{\bfq'})^2}
\right) \nn
&= 
\frac1{2\omega_{\bfq'}\omega^*_\bfq} 
\left(
\frac{\omega^*_\bfq+\omega_{\bfq'}-i\varepsilon}{(\omega^*_\bfq +\omega_{\bfq'}-i\varepsilon)^2-{P^0}^2}
+\frac{\omega^*_\bfq -\omega_{\bfq'}-i\varepsilon}{(\omega^*_\bfq -\omega_{\bfq'}-i\varepsilon)^2-{P^0}^2}
\right) \nn
&= 
\frac1{2\omega_{\bfq'}\omega^*_\bfq} 
\left(
\frac{\omega^*_\bfq+\omega_{\bfq'}}{(\omega^*_\bfq +\omega_{\bfq'})^2-i\varepsilon-{P^0}^2}
+\frac{\omega^*_\bfq -\omega_{\bfq'}}{(\omega^*_\bfq -\omega_{\bfq'})^2-{P^0}^2}
\right) \,.
\label{ConjugateGhost}
\end{align} 
Here we have used the fact that the 
$\delta_\rmc(P^0-q^0\pm\omega_{\bfq'})$  reduces to the usual Dirac delta 
$\delta(P^0-\Re q^0\pm\Re \omega_{\bfq'})$ for $q^0$ on the shifted contour 
$R(\pm\omega_\bfq)$, respectively. 
In going to the third equality, we have decomposed the preceding quantities
into four partial fractions by $(X^2-Y^2)^{-1}
=(2 X)^{-1}\sum_\pm (X\pm Y  )^{-1} $ and recombined them into  two
 terms.
In the last line, we have kept $i\varepsilon$ only in the 
denominator which may become 0.    

This result can be used to find the result for the ghost case 
$\phi_a=\varphi$ for which we also encounter no pole singularities 
in the shift of integration contour. 
The result can be obtained simply by replacing $\omega^*_\bfq-i\varepsilon$ 
in the above Eq.~(\ref{ConjugateGhost}) 
by $-(\omega_\bfq+i\varepsilon)$ since both poles $\omega^*_\bfq-i\varepsilon$ and 
$-(\omega_\bfq+i\varepsilon)$ are on the lower-half plane 
and placed infinitesimally below the contour $R(-\omega_{\bfq'})$. 
We find
\begin{align}
&\int_R dq^0\, \frac{1}{(\omega_\bfq+i\varepsilon)^2-{q^0}^2}\,
\cdot \sum_{\pm}\frac1{2\omega_{\bfq'}}
\delta_\rmc(P^0-q^0\pm\omega_{\bfq'}) \nn
&= 
-\frac1{2\omega_{\bfq'}\omega_\bfq} 
\left(
\frac{-\omega_\bfq+\omega_{\bfq'}-i\varepsilon}{(-\omega_\bfq +\omega_{\bfq'}-i\varepsilon)^2-{P^0}^2}
+\frac{-\omega_\bfq -\omega_{\bfq'}-i\varepsilon}{(-\omega_\bfq -\omega_{\bfq'}-i\varepsilon)^2-{P^0}^2}
\right) \nn
&= 
\frac1{4\omega_{\bfq'}\omega_\bfq} 
\left(
\frac1{\omega_\bfq -\omega_{\bfq'}+i\varepsilon+P^0}
+\frac1{\omega_\bfq -\omega_{\bfq'}+i\varepsilon-P^0}
+\frac{2(\omega_\bfq+\omega_{\bfq'})}{(\omega_\bfq +\omega_{\bfq'})^2-{P^0}^2}
\right)\\ 
&= 
\frac1{2\omega_\bfq^2} 
\left( -i\pi\delta(P^0) 
+\frac{2\omega_\bfq}{(2\omega_\bfq)^2-{P^0}^2}
\right) \,,
\label{Ghost}
\end{align} 
where we have set $\omega_\bfq=\omega_{\bfq'}$
in the CM frame in the last line.

Next is the `photon' term $\phi_a=A$ in the second line in Eq.
(\ref{eq:T-6}), which is similarly 
evaluated to this, 
aside from adding the `photon' pole contributions explained 
in (\ref{photonPole}). 
\begin{align}
&\int_R dq^0\, \frac{1}{{\nu_\bfq}^2-i\varepsilon-{q^0}^2}\,
\cdot \sum_{\pm}\frac1{2\omega_{\bfq'}}
\delta_\rmc(P^0-q^0\pm\omega_{\bfq'}) \nn
&=\int_{R(+\omega_\bfq)} dq^0\,  \frac{1}{{\nu_\bfq}^2-i\varepsilon-{q^0}^2}\,
\cdot \frac1{2\omega_{\bfq'}}
\delta_\rmc(P^0-q^0+\omega_{\bfq'}) 
+ \frac{2\pi i}{2\nu_\bfq\,2\omega_{\bfq'}}\delta_\rmc(P^0+\nu_\bfq+\omega_{\bfq'}) 
\nn
&\qquad {}+\int_{R(-\omega_\bfq)} dq^0\,  \frac{1}{{\nu_\bfq}^2-i\varepsilon-{q^0}^2}\,
\cdot \frac1{2\omega_{\bfq'}}
\delta_\rmc(P^0-q^0-\omega_{\bfq'}) 
+ \frac{2\pi i}{2\nu_\bfq\,2\omega_{\bfq'}}\delta_\rmc(P^0-\nu_\bfq-\omega_{\bfq'}) 
\nn
&= 
\frac{1}{2\omega_{\bfq'}\nu_\bfq} 
\biggl(
\frac{\nu_\bfq+\omega_{\bfq'}}{(\nu_\bfq +\omega_{\bfq'})^2-{P^0}^2}
+\frac{\nu_\bfq -\omega_{\bfq'}}{(\nu_\bfq -\omega_{\bfq'})^2-{P^0}^2} \nn
&\hspace{8em}{}+i\pi\delta_\rmc(P^0+\nu_\bfq+\omega_{\bfq'})
+i\pi\delta_\rmc(P^0-\nu_\bfq-\omega_{\bfq'})
\biggr) .
\label{Photon}
\end{align} 

Here we should add a comment on the fact that the above results 
(\ref{ConjugateGhost}) and (\ref{Ghost}) for the anti-ghost and ghost, 
respectively, are independent of the sign choice 
for the infinitesimal shift $+i\varepsilon$ or $-i\varepsilon$ of the pole position 
(\ref{PoleAvoid}). The first one (\ref{ConjugateGhost}) is the result for 
the anti-ghost with the pole shifted as $\omega^*_\bfq \rightarrow\omega^*_\bfq-i\varepsilon$, 
in which case the contour shift $R\ \rightarrow\ R(\pm\omega_\bfq)$ does not pick up 
those poles. Now, if we had chosen the opposite sign shift 
$\omega^*_\bfq \rightarrow\omega^*_\bfq+i\varepsilon$, then we meet the poles 
$\mp(\omega^*_\bfq+i\varepsilon)$ and would also have to pick up the contribution from the poles. 
The result for this case can immediately be obtained from the result (\ref{Photon}) 
for the photon by replacing $\nu_\bfq-i\varepsilon$ there by $\omega^*_\bfq+i\varepsilon$. This gives
\begin{align}
&\int_R dq^0\, \frac{1}{(\omega^*_\bfq+i\varepsilon)^2 -{q^0}^2}\,
\cdot \sum_{\pm}\frac1{2\omega_{\bfq'}}
\delta_\rmc(P^0-q^0\pm\omega_{\bfq'}) \nn
&= 
\frac1{2\omega_{\bfq'}\omega_\bfq^*} 
\biggl(
\frac{\omega^*_\bfq+\omega_{\bfq'}}{(\omega^*_\bfq +\omega_{\bfq'})^2+i\varepsilon-{P^0}^2}
+\frac{\omega^*_\bfq -\omega_{\bfq'}}{(\omega^*_\bfq -\omega_{\bfq'})^2-{P^0}^2} \nn
&\hspace{8em}{}+i\pi\delta_\rmc(P^0+\omega^*_\bfq+\omega_{\bfq'})
+i\pi\delta_\rmc(P^0-\omega^*_\bfq-\omega_{\bfq'})
\biggr) \,,
\label{Ghost*+ie}
\end{align} 
which is clearly equivalent to the previous result (\ref{ConjugateGhost}) 
because of the identities
\begin{align}
&\frac{\omega^*_\bfq+\omega_{\bfq'}}{(\omega^*_\bfq +\omega_{\bfq'})^2\pm i\varepsilon-{P^0}^2}
=
\frac12\left(\frac1{\omega^*_\bfq +\omega_{\bfq'}\pm i\varepsilon+P^0}
+\frac1{\omega^*_\bfq +c\pm i\varepsilon-P^0}
\right)\,, \nn
&\frac1{X -i\varepsilon} - \frac1{X +i\varepsilon} = 2\pi i \,\delta(X) \quad \hbox{for real $\forall X$}.
\label{eq:identity}
\end{align}
($(\omega^*_\bfq+\omega_{\bfq'})$ is real, because 
$\omega_{\bfq'}=\omega_{\bfq}$ in the CM frame.)
We can also see this equivalence for the ghost case; if we  adopted the 
shift $\omega_\bfq-i\varepsilon$, then we would have had to pick up the poles and should 
apply the `photon' result (\ref{Photon}) and replace $\nu_\bfq-i\varepsilon$ there 
by $-(\omega_\bfq-i\varepsilon)$. It would give 
\begin{align}
&\frac1{2\omega_{\bfq'}(-\omega_\bfq)} 
\biggl(
\frac12\Big(\frac1{-\omega_\bfq +\omega_{\bfq'}+i\varepsilon-P^0}
+\frac1{-\omega_\bfq +\omega_{\bfq'}+i\varepsilon+P^0}\Big)
+\frac{-\omega_\bfq -\omega_{\bfq'}}{(-\omega_\bfq -\omega_{\bfq'})^2-{P^0}^2} \nn
&\hspace{8em}{}+i\pi\delta_\rmc(P^0-\omega_\bfq+\omega_{\bfq'})
+i\pi\delta_\rmc(P^0+\omega_\bfq-\omega_{\bfq'})
\biggr) \nn
&\stackrel{\bfq'\ \rightarrow\ -\bfq}{\longrightarrow}\ \ 
-\frac1{2\omega_\bfq^2} 
\biggl(\frac12\Big(
\frac1{+i\varepsilon-P^0}
+\frac1{+i\varepsilon+P^0}\Big)
-\frac{2\omega_\bfq}{(2\omega_\bfq)^2-{P^0}^2} + 2\pi i\,\delta(P^0)
\biggr)\,,
\end{align} 
which is equal to the previous result (\ref{Ghost}) owing to the 
same identity (\ref{eq:identity}).

We summarize the results:
The first line integral in the scattering $T$-amplitude (\ref{eq:T-6}) is 
given by Eq.~(\ref{FirstLine}):
\begin{align}
&f^2\int\frac{d^3\bfq}{(2\pi)^3} \sum_{a.b}\eta_a\eta_b 
\frac1{2\omega^{La}_\bfq\, \omega^{Lb}_{\bfq}} \left(
\frac{\omega^{Lb}_{\bfq}+\omega^{La}_\bfq}{(\omega^{Lb}_{\bfq}+\omega^{La}_\bfq)^2-{P^0}^2}
\right) \nn
&=
f^2\int\frac{d^3\bfq}{(2\pi)^3} \times\ \ 
\left\{\begin{array}{ccc}
 \phi_a, \phi_b  &  \text{comb.}\eta_a\eta_b  &        \\ \hline
A,A :  &  1    
& \disp \frac1{2\nu_\bfq^2}\,\frac{2\nu_\bfq}{(2\nu_\bfq)^2-{P^0}^2-i\varepsilon} \\
A,\varphi: & 2\cdot \left(-\frac12\right)  
& \disp  \frac1{2\nu_\bfq\, (-\omega_\bfq)}\,\frac{\nu_\bfq-\omega_\bfq}{(\nu_\bfq-\omega_\bfq)^2-{P^0}^2} \\      
A,\varphi^\dagger: & 2\cdot \left(-\frac12\right)  
& \disp \frac1{2\nu_\bfq\, \omega^*_\bfq}\,\frac{\nu_\bfq+\omega^*_\bfq}{(\nu_\bfq+\omega^*_\bfq)^2-{P^0}^2} \\      
\varphi, \varphi^\dagger: & 2\cdot \left(-\frac12\right)^2  
& \disp \frac1{2(-\omega_\bfq)\omega^*_\bfq}\,\frac{-\omega_\bfq+\omega^*_\bfq}{(-\omega_\bfq+\omega^*_\bfq)^2-{P^0}^2} \\      
\varphi, \varphi: &  \left(-\frac12\right)^2  
& \disp \frac1{2(-\omega_\bfq)^2}\,\frac{-2\omega_\bfq}{(-2\omega_\bfq)^2-{P^0}^2} \\      
\varphi^\dagger, \varphi^\dagger: & \cdot \left(-\frac12\right)^2  
& \disp \frac1{2(\omega^*_\bfq)^2}\,\frac{2\omega^*_\bfq}{(2\omega^*_\bfq)^2-{P^0}^2} \\      
\end{array}\right.
\label{eq:1st}
\end{align}
The second line in Eq.~(\ref{eq:T-6}) is given in 
Eq.~(\ref{Photon}) for `photon' case $\phi_a=A$, 
Eq.~(\ref{ConjugateGhost}) for anti-ghost case $\phi_a=\varphi^\dagger$ and  
Eq.~(\ref{Ghost}) for ghost case $\phi_a=\varphi$, respectively:  
\begin{align}
&f^2 \int\frac{d^3\bfq}{(2\pi)^3}
\sum_{a} 2\eta_a \eta_\varphi 
\int_R dq^0 \frac1{{\omega^a_\bfq}^2-{q^0}^2}\,
\cdot \sum_{\pm}\frac1{2\omega_{\bfq}}
\delta_\rmc(P^0-q^0\pm\omega_{\bfq}) \nn
&=
f^2\int\frac{d^3\bfq}{(2\pi)^3} \times 
\left\{ \begin{array}{ccl}
 \phi_a  &  2\eta_a\eta_\varphi  &        \\ \hline
A :  &  2 \left(-\frac12\right)    
& \disp 
\frac1{2\omega_\bfq \nu_\bfq} 
\biggl(
\frac{\nu_\bfq+\omega_{\bfq}}{(\nu_\bfq +\omega_{\bfq})^2-{P^0}^2}
+\frac{\nu_\bfq -\omega_{\bfq}}{(\nu_\bfq -\omega_{\bfq})^2-{P^0}^2}  \\
& & \disp \hspace{3.4em}{}+i\pi\delta_\rmc(P^0+\nu_\bfq+\omega_{\bfq})
+i\pi\delta_\rmc(P^0-\nu_\bfq-\omega_{\bfq})
\biggr) \\.
\varphi^\dagger: & 2 \left(-\frac12\right)^2
& \disp  
\frac1{2\omega_{\bfq}\omega^*_\bfq} 
\left(
\frac{\omega^*_\bfq+\omega_{\bfq}}{(\omega^*_\bfq +\omega_{\bfq})^2-i\varepsilon-{P^0}^2}
+\frac{\omega^*_\bfq -\omega_{\bfq}}{(\omega^*_\bfq -\omega_{\bfq})^2-{P^0}^2}
\right) \\
\varphi: & 2 \left(-\frac12\right)^2
& \disp 
\frac1{2\omega_\bfq^2} 
\left( -i\pi\delta(P^0) 
+\frac{2\omega_\bfq}{(2\omega_\bfq)^2-{P^0}^2}
\right) 
\end{array}\right.
\label{eq:2nd}
\end{align}

[III] {\it Finally the third line in (\ref{eq:T-6})} reads 
\begin{equation}
f^2 \int\frac{d^3\bfq}{(2\pi)^3}
\left(-\frac12\right)^2\frac{2\pi i}{(2\omega_\bfq)^2}
\left(
\delta_\rmc(P^0+2\omega_\bfq) + 2\delta(P^0) + \delta_\rmc(P^0-2\omega_\bfq)
\right)\,.
\label{eq:3rd}
\end{equation}
We note that the terms containing the energy differences $\nu_\bfq-\omega_\bfq$ 
or $\omega^*_\bfq-\omega_\bfq$ in the numerator are exactly canceled between 
Eqs.~(\ref{eq:1st}) and (\ref{eq:2nd}), and the second last term of    
Eq.~(\ref{eq:1st}) cancels the half of the last term in Eq.~(\ref{eq:2nd}).
The $\delta(P^0)$ terms   in Eqs.~(\ref{eq:2nd}) and (\ref{eq:3rd}) also cancel.
All the rest terms in these Eqs.~(\ref{eq:1st}) to (\ref{eq:3rd}) 
exactly coincide term by term with those in Eq.~(\ref{eq:S-1-5}) obtained 
by using 3d momentum form of propagators.

\bibliographystyle{JHEP}

\bibliography{library}

\end{document}